\pdfoutput=1
\documentclass[aip,pop,reprint,nofootinbib]{revtex4-1}
\usepackage{lmodern}
\usepackage[T1]{fontenc}
\usepackage{graphicx}
\usepackage{bm,amssymb}
\usepackage{xfrac}
\usepackage{xcolor}
\usepackage{mathtools}
\usepackage{array,multirow}
\usepackage{float}
\usepackage{cancel}
\usepackage{tabularx}
\usepackage{float}
\usepackage{tikz}
\usetikzlibrary{calc,patterns,fadings,intersections}
\usepackage{zref-savepos}
\usepackage{soul}
\usepackage{microtype}
\newcounter{DiagonalizedEntry}
\usepackage{upgreek}
\usepackage{siunitx}

\usepackage{pgfplots}
\usetikzlibrary{calc}
\pgfplotsset{compat=1.14}

\definecolor{themeRed}{RGB}{146,0,0}
\definecolor{themeGreen}{RGB}{0,146,146}
\definecolor{themeOrange}{RGB}{219,109,0}
\definecolor{themePurple}{RGB}{73,0,146}
\definecolor{themeGray}{RGB}{146,146,146}
\definecolor{themePink}{RGB}{255,109,182}
\definecolor{themeBlue}{RGB}{109,182,255}

\pgfplotsset{local/.style={black,dashed}}
\pgfplotsset{vfp/.style={themeGreen}}
\pgfplotsset{snb/.style={themeOrange,densely dotted}}
\pgfplotsset{fluxlim/.style={themeRed,dashdotted}}
\pgfplotsset{eic/.style={themeRed,dashdotted}}
\pgfplotsset{nflf/.style={themePurple,dashed}}

\tikzset{
every picture/.append style={baseline,trim left={($(current axis.south west)-(1.cm,0cm)$)},trim axis right}
}
\pgfplotsset{
every axis legend/.append style={draw=none,fill=none,legend cell align=left,cells={align=left},font=\footnotesize},
every axis/.append style={scale only axis,width=0.8\linewidth,height=0.6\linewidth,title={\phantom{moo}},title style={at={(0.5,.96)}}},
every axis plot/.append style={thick},
every y tick label/.append style={font=\footnotesize,
/pgf/number format/fixed},
every x tick label/.append style={font=\footnotesize,/pgf/number format/fixed},
}

\usepgfplotslibrary{external}
\tikzset{external/force remake=false}
\expandafter\gdef\csname c@tikzext@no@\pgfkeysvalueof{/tikz/external/figure name}\endcsname{1}%

\renewcommand*{\theDiagonalizedEntry}{NTE-\the\value{DiagonalizedEntry}}

\newcommand*{\citen}{}
\DeclareRobustCommand*{\citen}[1]{%
  \begingroup
    \romannumeral-`\x 
    \setcitestyle{numbers}%
    [\cite{#1}]%
  \endgroup
}

\usepackage[colorlinks,
    linkcolor={blue!80!black},
    citecolor={blue!80!black},
    urlcolor={blue!80!black}]{hyperref}
    
\begin{document}


\title{Testing \textcolor{black}{nonlocal models of electron thermal conduction} for magnetic and inertial confinement fusion applications}


\author{J.~P.~Brodrick}
\email{jonathan.brodrick@york.ac.uk}
\affiliation{York Plasma Institute, Dep't of Physics, University of York, Heslington, York YO10 5DD, United Kingdom}
\author{R.~J.~Kingham}
\affiliation{Plasma  Physics Group, Blackett Laboratory, Imperial College, London SW7 2BW, United Kingdom}\author{M.~M.~Marinak}
\author{M.~V.~Patel}
\affiliation{Lawrence Livermore National Laboratory, 7000 East Ave., Livermore, CA, USA}
\author{A.~V.~Chankin}
\affiliation{Max-Planck-Institute for Plasma Physics, Boltzmannstr. 2, 85748 Garching, Germany}
\author{J.~T.~Omotani}
\affiliation{Dep't of Physics, Chalmers University of Technology, SE-412 96 Gothenburg, Sweden}
\author{M.~V.~Umansky}
\affiliation{Lawrence Livermore National Laboratory, 7000 East Ave., Livermore, CA, USA}
\author{D.~Del~Sorbo} 
\author{B.~Dudson}
\affiliation{York Plasma Institute, Dep't of Physics, University of York, Heslington, York YO10 5DD, United Kingdom}
\author{J.~T.~Parker}
\affiliation{Science Technology and Facilities Council, Rutherford Appleton Laboratory, Harwell Campus, Didcot OX11 0QX, United Kingdom}
\author{G.~D.~Kerbel}
\author{M.~Sherlock}
\affiliation{Lawrence Livermore National Laboratory, 7000 East Ave., Livermore, CA, USA}
\author{C.~P.~Ridgers}
\affiliation{York Plasma Institute, Dep't of Physics, University of York, Heslington, York YO10 5DD, United Kingdom}

\date{\today}

\begin{abstract}
Three models for nonlocal \textcolor{black}{electron thermal} transport are here compared against Vlasov-Fokker-Planck (VFP) codes to assess their accuracy in situations relevant to both inertial fusion hohlraums and tokamak scrape-off layers. The models tested are (i) a moment-based approach using an eigenvector integral closure (EIC) originally developed by Ji, Held and Sovinec; (ii) the non-Fourier Landau-fluid (NFLF) model of Dimits, Joseph and Umansky; and (iii) Schurtz, Nicola\"{i} and Busquet's multigroup diffusion model (SNB). We find that while the EIC and NFLF models accurately predict the damping rate of a small-amplitude temperature perturbation (within $10\%$ at moderate collisionalities), they overestimate the peak heat flow by as much as 35\%  and do not predict preheat in \textcolor{black}{the more relevant} case where there is a large temperature difference.  The SNB  model, however, agrees better with VFP results for the latter problem \textcolor{black}{if care is taken with the definition of the mean free path}.  Additionally, we present for the first time a comparison of the SNB model against a VFP code for a hohlraum-relevant problem with inhomogeneous ionisation and show that the model overestimates the heat flow in the helium gas-fill by a factor of $\mathord{\sim}2$ despite predicting the peak heat flux to within  16\%.
\end{abstract}


\maketitle


\section{Introduction}

Performing full integrated simulations of large-scale fusion devices, such as the National Ignition Facility (NIF) or the ITER tokamak, is very challenging due to the wide range of scales over which the important physical processes occur. Codes, such as HYDRA\cite{HYDRA} and BOUT++\cite{BOUTframework}, used to perform integrated simulations of inertial and magnetic confinement fusion (ICF/MCF) respectively, often include reduced models to capture complex aspects of the physics. The accuracy of the models used naturally affects the predictive capability of these codes. In this paper we compare three different models for kinetic (i.e. nonlocal) effects on \textcolor{black}{electron} thermal \textcolor{black}{conduction against Vlasov-Fokker-Planck simulations}: \textcolor{black}{ (i) the EIC\cite{JiHeld,Omotani,Hermite} and (ii) the NFLF\cite{NFLFTheory,*NFLF2014,NFLFImplementation} models, which have recently been suggested for application in the tokamak edge and scrape-off layer (SOL); and (iii) the SNB model\cite{SNB,SNBmag,Dario,Dariomag,Cao}, which is currently the most widely used in inertial fusion and laser-plasma applications.}

\textcolor{black}{In collisional plasmas (where \textcolor{black}{the mean free path is much less than the temperature scalelength)}, the electron heat flow, parallel to any macroscopic magnetic field in the plasma, has been shown by Braginskii\cite{Braginskii} to obey Fourier's law}:
\begin{equation}
\label{braginskii_q}
\bm{Q}^\mathrm{(B)} = -\kappa^\mathrm{(B)} n_\mathrm{e} v_\mathrm{T} \lambda_\mathrm{ei}^{\textrm{(B)}} \nabla k_\mathrm{B} T_\mathrm{e},
\end{equation}
\textcolor{black}{\noindent where $\kappa^\mathrm{(B)}$ is the dimensionless thermal conductivity, $n_\mathrm{e}$ the electron density,  $v_\mathrm{T} = \sqrt{k_\mathrm{B} T_\mathrm{e}/m_\mathrm{e}}$ is the thermal velocity, 
\begin{equation}
\lambda_\mathrm{ei}^{\textrm{(B)}} =  3\sqrt{\frac{\uppi}{2}} \frac{\left(k_\mathrm{B} T_\mathrm{e}\right)^2}{4\uppi Zn_\mathrm{e} e^4\log\Lambda_\mathrm{ei}}
\end{equation}
\noindent is an averaged electron-ion mean free path (mfp) in Gaussian units (which shall be used throughout this paper),  $k_\mathrm{B}$ is Boltzmann's constant,  $T_\mathrm{e}$ is the electron temperature, $Z$ is the average ionisation, $e$ is the magnitude of the electron charge,  and $\log\Lambda_\mathrm{ei}$ is the Coulomb logarithm for electron-ion scattering which is typically between 2 and 10 in cases of interest here.} Here and for the entirety of this paper we assume there to be no drift velocity or current (hence the ion and electron rest frames are equivalent). Note that Epperlein and Haines\cite{EppHaines} have calculated $\kappa^\mathrm{(B)}$ to an increased accuracy and Epperlein and Short\cite{EppShort} later suggested that this can be approximated well by $\kappa^\mathrm{(B)} \approx \xi(Z) 128/3\uppi$, where $\xi(Z) = (Z+0.24)/(Z+4.2)$.

Equation \ref{braginskii_q} breaks down if the collisionality of the electrons becomes low.  This is due to the inadequacy of the assumption that the electron distribution function is close to Maxwellian; electrons with mfp's larger than the temperature scalelength can in fact escape gradients before being scattered and depositing their energy into the plasma, \textcolor{black}{leading to a distortion of the distribution function away from Maxwellian.}

\textcolor{black}{The largest contribution to the heat flow comes from suprathermal electrons with a velocity of approximately $4v_\mathrm{T}$. Due to the $v^4$ scaling of the appropriate mfp's, these suprathermals can travel over \textcolor{black}{a hundred} times further than thermal electrons enabling} excess heat to be deposited beyond the steepest part of the temperature profile (often referred to as `preheat' in the literature \cite{EppShort}).  A reduced population of suprathermals is left behind in the region of steep temperature gradient, \textcolor{black}{contributing to a reduction in} the heat flux.  These `nonlocal' effects can become important even for temperature scalelengths as long as $\mathord{\sim}200$ thermal mfp's.\cite{SNB}

Such situations occur frequently in \textcolor{black}{important regions of both MCF and ICF experiments}: In tokamaks, nonlocal thermal transport is thought to play a role in heat flow from the core plasma to the `divertor'\cite{Chodura}, a region of the tokamak edge specifically designed to absorb and exhaust excess heat from the plasma. Thermal electrons entering the SOL at the separatrix have mfp's ranging from 1\% (C-Mod) to 20\% (DIII-D/\textcolor{black}{Tokamak de Varennes} (TdeV)) of the distance to the divertor target (connection length). \textcolor{black}{For ITER this is estimated to be 8\%.} In fact, the ratio of $\lambda_\mathrm{ei}^{\textrm{(B)}}$ to the local temperature scalelength $L_T = 1/\nabla_\parallel \log T_\mathrm{e}$ tends to vary along the SOL from approximately 1 (TdeV) or 0.1 (DIII-D) near the separatrix, to 0.01 near the colder divertor.\cite{Batishchev} \textcolor{black}{These ratios are all two orders of magnitude higher for suprathermal electrons,} rendering the heat transport inherently nonlocal. Furthermore, transient events---Edge Localised Modes (ELMs), disruptions and filaments---can create even higher temperatures and steeper gradients, with which the associated suprathermals would be almost collisionless.\cite{Omotani}

Current state of the art codes, such as SOLPS\cite{SOLPS1,SOLPS2} and UEDGE\cite{UEDGE}, have been shown to significantly underestimate the outer divertor target electron temperature and overestimate its density compared to experiment in existing tokamaks, which in turn affects other plasma parameters. Chankin and Coster \cite{Chankin2D} have suggested that nonlocal effects in addition to \textcolor{black}{inadequate} treatment of neutrals (which has largely been ruled out by a sensitivity analysis) and inappropriate use of time-averaging could explain this discrepancy. The plausibility of this hypothesis is supported by recent gyrokinetic simulations performed by Churchill \emph{et al.}\cite{Churchill}. \textcolor{black}{Another important factor is the effect of the enhanced suprathermal population on Langmuir probe characteristics\cite{Horacek2003,JAWORSKI2012,JAWORSKI2013,Izacard}: \v{D}uran \emph{et al.}\cite{Duran2015} have shown that a more sophisticated interpretation of probe results can reduce but not eliminate the discrepancy.} Resolution of this discrepancy is critical as excessive heat loads could erode and severely limit the lifetimes of the divertor target plates.\cite{Turnyanskiy}


\textcolor{black}{For the case of indirect-drive inertial fusion, steep temperature gradients of approximately $100\;\mathrm{\upmu m}$ are set up near the inner surface of the gold hohlraum that contains both the helium gas-fill and the fuel capsule. This is induced by the high-energy lasers which ionise and ablate the hohlraum wall. Ratios of $\lambda_\mathrm{ei}^{\textrm{(B)}}/L_T$ exceeding 10-20\% in this region have been reported.\cite{SNB} Significant nonlocal effects on the thermal conduction are consequently observed, particularly in the neighbouring low-density gas-fill where the mfp's of heat-carrying electrons can be very long.  Failure to simulate this nonlocality accurately can have implications for hohlraum temperatures and implosion symmetry.\cite{HYDRA}}

A common approach to incorporate the flux reduction aspect of nonlocal transport is to restrict the local heat flow to some fraction $\mathfrak{f}_\mathrm{L}$ of the free-streaming limit $Q_\mathrm{fs} =  n_\mathrm{e} k_\mathrm{B} T_\mathrm{e} v_\mathrm{T}$. However, at best the flux-limiter $\mathfrak{f}_\mathrm{L}$ must be tuned against previous experiments, limiting predictive capability---values ranging from 0.03 to \textcolor{black}{ 0.15 have been suggested for NIF design codes \cite{HYDRA,Jones} and up to 3} for SOL modeling\cite{fundamenski}---and preheat effects cannot be predicted.

A more complete way to take nonlocal effects into account, however, is with a fully kinetic approach. By solving the Vlasov-Fokker-Planck (VFP) equation for the electron distribution function directly \textcolor{black}{(along with self-consistent electric and magnetic fields)} we need not assume \textcolor{black}{ it is close to} Maxwellian; nonlocal effects are calculated ab-initio. Such an approach typically assumes binary collisions and small-angle scattering limiting its applicability in regions where the plasma is strongly coupled ($\log \Lambda$ approaching unity) such as in ICF fuel capsules or the colder part of the partially ionised hohlraum wall. While it is possible for VFP codes to treat collisions between multiple ion species\cite{IFP0D2V} and even neutrals\cite{VFPNeutrals} (though the latter might require coupling to a neutral Monte Carlo code such as EIRENE\cite{KIPPDev,KIPPBenchmarks,B2Eirene} due to the importance of large-angle collisions) here we only consider collisions of electrons with a single stationary ion species. Quantum-mechanical effects such as Fermi degeneracy which could be of importance in solid density material are also typically negelcted\cite{VFPReview}, nevertheless methods to incorporate these have been suggested\cite{FermiDirac}

However, due to the extra dimensionality associated with solving in velocity-space, VFP codes are computationally intensive and difficult to incorporate into existing integrated modeling codes. Such demands of resolving the distribution function and collision time are especially restrictive in cold, dense regions such as deep in the hohlraum wall where a fluid treatment might even be sufficient. Therefore, alternative models that have more predictive capability than flux-limiters, and reduced computational requirements compared to a full kinetic simulation, are desirable. A dedicated experiment to measure nonlocal effects performed by Gregori \emph{et al.}\cite{Gregori} has \textcolor{black}{shown that a model of this kind can approximate measured temperature profiles better than flux-limiters}.\color{black}

\color{black}
A large number of reduced models for nonlocal electron thermal transport have been suggested for applications in inertial fusion and laser-plasmas \cite{EppShort,LMV,CMG,SNB,Batischev2002,SNBmag,Brantov2013,Dario,Dariomag,Cao} and to the SOL \cite{Hammett,JiHeld,Omotani,NFLFTheory,*NFLF2014,Hermite,NFLFImplementation,Izacard,Izacard2017}. This paper focuses on three of these models, here referred to as the SNB\cite{SNB,SNBmag,Dario,Dariomag,Cao}, EIC\cite{JiHeld,Omotani,Hermite} and NFLF\cite{NFLFTheory,*NFLF2014,NFLFImplementation} models \textcolor{black}{(described in section \ref{sec:models})}, and compares them for accuracy against VFP simulations.While the SNB model has recently been compared to VFP results by Marocchino \emph{et al.}\cite{Marocchino}, this has shown that the two approaches agree well for $Z=1$ but begin to deviate from each other as the ionisation increases. This is surprising as the SNB model was originally derived in the high-$Z$ (Lorentz) limit and we observe here that such findings are sensitive to precise implementation details of the model. Additionally, while the EIC and NFLF models have been shown to predict similar heat-fluxes\cite{NFLFImplementation}, they have not yet been validated against a full time-dependent VFP code.

The first problem we investigate here is the damping of a small-amplitude sinusoidal temperature profile of various wavelengths in section \ref{sec:linear}. This test will be used to justify a tuning parameter which can be applied to the SNB model to improve agreement with VFP simulations. We will additionally suggest a new analytic fit for the conductivity reduction and use this to obtain improved coefficients for the NFLF model.

In section \ref{sec:homo}, we will then consider the case, more relevant \textcolor{black}{to both hohlraums and the SOL}, of a plasma with a large temperature variation.  We will show that the SNB model agrees very well with VFP simulations using the same tuning factor as in the linearised problem described above and that the EIC and NFLF models overpredict the peak heat flux.  While this suggests that the SNB model may also be useful in SOL simulations, we also consider potential improvements to the other models to improve their performance.

Finally, we will show in section \ref{sec:varion} that the SNB model can significantly overpredict the heat flow relative to VFP in the low-density helium gas-fill using a problem particularly relevant to the \textcolor{black}{ablated} hohlraum wall. The importance of gradients in both average ionisation $Z$ and electron density $n_\mathrm{e}$ here could mean the findings could also be important for the detached divertor scenario.

\section{Vlasov-Fokker-Planck Modeling}  \label{sec:vfp}

The evolution of the \textcolor{black}{electron} distribution function $f_\mathrm{e}$, \textcolor{black}{assuming \textcolor{black}{small-angle scattering from} binary collisions}, can be \textcolor{black}{described} by the Vlasov-Fokker-Planck equation\cite{IMPACT}
\begin{equation}
\frac{\partial f_\mathrm{e}}{\partial t} + \bm{v}\cdot\nabla f_\mathrm{e}- \frac{e}{m_\mathrm{e}}\left(\bm{E}+\frac{\bm{v}\times\bm{B}}{c}\right)\cdot \frac{\partial f_\mathrm{e}}{\partial \bm{v}} = C(f_\mathrm{e}),
\end{equation}
where $\bm{v}$ is the electron velocity, $\bm{E}$, $\bm{B}$ are the electric and magnetic fields  respectively, $m_\mathrm{e}$ is the electron mass. \textcolor{black}{Two of the three VFP codes used here, IMPACT and KIPP, employ a zero-current constraint, \textcolor{black}{$\int f \bm{v} \;\mathrm{d}^3\bm{v}=\nobreak 0$} to calculate the electric field, which ensures quasineutrality. The third VFP code, SPRING, uses a more sophisticated approach which solves the Poisson and ion continuity equations with an implicit-moment method\cite{Epperlein94,implicitPoisson}.}

In this work we consider only collisions of electrons with themselves and a single ion species using the standard Trubnikov-Rosenbluth\cite{Trubnikov,Rosenbluth} form of the Fokker-Planck collision operator $C=C_\mathrm{ee}+C_\mathrm{ei}$, where
\begin{equation}
\begin{split}
\frac{C_{\mathrm{e}\beta}(f_\mathrm{e},f_\beta)}{\Gamma_{\mathrm{e}\beta}}=&-\frac{\partial}{\partial \bm{v}_i}
\Bigg(\frac{m_\mathrm{e}}{m_\beta}f_\mathrm{e}\frac{\partial}{\partial \bm{v}_i} \int \frac{f_\beta}{\lvert \bm{v}-\bm{u}\rvert}\; \mathrm{d}^3 \bm{u}\\
&-\frac{1}{2}\frac{\partial f_\mathrm{e}}{\partial \bm{v}_j}\frac{\partial^2 }{\partial \bm{v}_i\partial\bm{v}_j}\int {f_\beta}\lvert \bm{v}-\bm{u}\rvert\; \mathrm{d}^3 \bm{u}\Bigg)
\end{split}
\end{equation}
(applying standard Einstein summation over repeated indices). Here we have defined
\begin{equation}
\Gamma_{\mathrm{e}\beta} = \frac{4\uppi Z_\mathrm{e}^2Z_\beta^2e^4}{m_\mathrm{e}^2}\log \Lambda_{\mathrm{e}\beta},
\end{equation}
where $Z_\mathrm{i}=Z$ is the average ionisation and $Z_\mathrm{e}=-1$, along with $m_\mathrm{i}$ the ion mass. The ion distribution function $f_\mathrm{i}$ is assumed by KIPP to be Maxwellian, here we enforce the ion temperature to be equal to the electron temperature but this is not necessary\cite{KIPPRelax}, all other codes and models assume a cold ion population and neglect terms of order $m_\mathrm{e}/m_\mathrm{i}$, simplifying the electron-ion collision operator to:
\begin{equation}
\frac{C_\mathrm{ei}(f,n_\mathrm{i}\delta(\bm{v}))}{n_\mathrm{i}\Gamma_{\mathrm{ei}}}=\frac{\partial}{\partial \bm{v}_i}
\Bigg(\cancelto{0}{\frac{m_\mathrm{e}}{m_\mathrm{i}}\frac{\bm{v}_i}{v^3}f_\mathrm{e}} +\frac{v^2\delta_{ij}-\bm{v}_i\bm{v}_j}{2v^3}\frac{\partial f_\mathrm{e}}{\partial \bm{v}_j}\Bigg),
\end{equation}
where $\delta(\bm{v})$ is the Dirac delta function and $\delta_{ij}$ is the Kronecker delta tensor.

For the case where variations only occur along magnetic field lines, symmetry in the perpendicular direction allows for elimination of the magnetic field by `gyro-averaging' (\textcolor{black}{integrating azimuthally around the $v_\parallel$ axis,} this process is still valid even in the absence of magnetic fields); this yields the 1d2v \textcolor{black}{(one-dimensional in space, two-dimensional in velocity)} equation
\begin{equation}
\frac{\partial \langle f_\mathrm{e}\rangle}{\partial t} + v_\parallel\frac{ \partial\langle f_\mathrm{e} \rangle}{\partial s_\parallel} - \frac{e E_\parallel}{m_\mathrm{e}} \frac{ \partial \langle f_\mathrm{e}\rangle}{\partial v_\parallel} = \langle C(f_\mathrm{e})\rangle,
\end{equation}
where $\langle\cdot\rangle$ represents a gyro-average (an explicit representation of $\langle C\rangle$ can be found in previous work by Xiong \emph{et al.}\cite{Xiong} and Chankin \emph{et al.}\cite{KIPPDev}), and $\parallel$ denotes components of vectors parallel to the magnetic field.

The KIPP code\cite{KIPPDev} is designed to solve this equation using Cartesian spatial and velocity grids. The code uses an operator splitting method suggested by Shoucri and Gagne\cite{opSplit,ShoucriBook} that treats the spatial derivative using a second-order explicit scheme followed by the electric field and collision operator terms using a first-order (in time, second-order in velocity) implicit scheme. The velocity grid covers the domain $  v_\parallel \in [-v_\mathrm{max}, v_\mathrm{max}],  v_\perp \in [0, v_\mathrm{max}]$ where $v_\mathrm{max}$ is a user-defined parameter. The distribution function is simply taken to be zero at the exterior of the grid and reflected along the interior $v_\perp=0$ axis.

A simplified approach is the diffusion approximation, which consists of expanding the distribution function in Cartesian tensors and truncating all but the first two terms ($f_\mathrm{e}=f_0(v)+\bm{v}\cdot\bm{f}_1(v)/v$). \textcolor{black}{This reduces the number of velocity-space dimensions to one thereby} increasing efficiency and has been observed to correctly predict heat flows to within 5\% for temperature scalelengths $L_T \lessapprox 10\lambda_\mathrm{ei}^\mathrm{(B)}$\cite{MatteVirmont}. The IMPACT code\cite{IMPACT}  \textcolor{black}{(\textcolor{black}{two-dimensional} in space)} employs this approach and makes a further simplification of ignoring angular scattering due to electron-electron collisions, valid in the Lorentz limit. In order to recover the correct local thermal conductivity for lower-$Z$ plasmas the factor $\xi(Z)$ is used in the electron-ion collision frequency. Our comparisons between IMPACT and KIPP suggest that these approximations do not greatly affect the results for the problems studied in section \ref{sec:homo}. \textcolor{black}{The equations solved by IMPACT, along with Ampere and Faraday's Law are thus}
\begin{align}
\label{eq:f0_orig}
\frac{\partial f_0}{\partial t}+\frac{v}{3}\nabla \cdot \bm{f}_1  - \frac{e\bm{E}}{3m_\mathrm{e}v^2}\cdot\frac{\partial v^2 \bm{f}_1}{\partial v} &= C_{\mathrm{ee0}}\left[f_0\right], \\
\label{eq:f1_orig}
\frac{\partial \bm{f}_1}{\partial t}+v\nabla f_0 - \frac{e\bm{E}}{m_\mathrm{e}}\frac{\partial f_0}{\partial v}-\frac{e\left(\bm{B}\times\bm{f}_1\right) }{m_\mathrm{e} c}&= - \frac{\nu_\mathrm{ei}}{\xi}\bm{f}_1,
\end{align}
where
\begin{equation}
\begin{split}
C_\mathrm{ee0}[f_0] = \frac{4\uppi \Gamma_\mathrm{ee}}{v^2}\frac{\partial}{\partial v}\bigg(\int_{\mathrlap{0}}^{\mathrlap{v}} f_0 u^2\; &\mathrm{d}u f_0\\
 + \frac{1}{v}\int_{\mathrlap{0}}^{\mathrlap{v}} u^2 \int_{\mathrlap{u}}^{\mathrlap{\infty}} f_0 w \;&\mathrm{d} w \;\mathrm{d}u\frac{\partial f_0}{\partial v}\bigg)
\end{split}
\end{equation}
is the isotropic contribution of the electron-electron collision operator and
\begin{equation}
\nu_\mathrm{ei} = \frac{n_\mathrm{i}\Gamma_\mathrm{ei}}{v^3}=\frac{4\uppi Zn_\mathrm{e} e^4 \log\Lambda_\mathrm{ei}}{m_\mathrm{e}^2v^3}
\end{equation}
 is the velocity-dependent electron-ion collision frequency.

IMPACT is fully implicit and first order in time, and uses fixed-point/Picard iterations to handle nonlinear terms. Note that the implicit treatment of the electric field \textcolor{black}{enforces} charge conservation at every iteration. \textcolor{black}{The magnetic field and electron inertia (${\partial \bm{f}_1}/{\partial t}$) terms have not been included in the simulations appearing in this paper.} The main reason for using IMPACT in section \ref{sec:varion} is that KIPP has not yet been extended to spatially-varying ionisation along $s_\parallel$.

Finally, we also include results  previously obtained with the SPRING\cite{Epperlein94} VFP code which takes a Cartesian tensor expansion to arbitrary order and does not neglect/approximate \textcolor{black}{anisotropic} electron-electron collisions. This code uses a linearised approach, i.e. the spatial gradient operator $\nabla$ is replaced by $\mathrm{i}k$, making it advantageous for analysing the \textcolor{black}{small-amplitude sinusoidal temperature perturbations} featured in section \ref{sec:linear}, but not problems with large temperature perturbations.
\color{black}
\section{Nonlocal models}\label{sec:models}

\subsection{Eigenvector Integral Closure}

The first model investigated here was originally proposed by Ji, Held and Sovinec \cite{JiHeld} and is directly derived from simplifications of the VFP equation. Necessarily, the time-derivative term is \textcolor{black}{neglected} to allow the heat flow to be calculated based on input density and temperature profiles only, and not the history of the distribution function; this assumption  \color{black} should be reasonable over `mean' SOL profiles (i.e. averaged over time to eliminate fine-scalelength fluctuations), \color{black}but could break down for transient events with faster timescales such as filaments and ELMs. 

The distribution function is expressed as $f_\mathrm{e} = f^{(0)} + \delta f$, where $\delta f$ is a perturbation to an initial, Maxwellian, guess $f^{(0)}$. Assuming the perturbation $\delta f$ is small, the nonlinear collision and electric field terms in the gyro-averaged VFP equation are linearised, which yields
\begin{equation}
\label{eq:DKE}
\frac{\partial \langle\delta f\rangle}{\partial s_\parallel} - \frac{\langle C_\mathrm{L}(\delta f)\rangle}{v_\parallel}  = \frac{e E_\parallel}{m_\mathrm{e}v_\parallel}\frac{\langle\partial f^{(0)}\rangle}{\partial v_\parallel}-\frac{\partial \langle f^{(0)}\rangle}{\partial s_\parallel},
\end{equation}
where 
\begin{equation}
C_\mathrm{L}(\delta f) =  C_\mathrm{ee}(f^\mathrm{(0)},\delta f)+  C_\mathrm{ee}(\delta f,f^\mathrm{(0)})+C_\mathrm{ei}(\delta f,n_\mathrm{i}\delta(\bm{v}))
\end{equation}
 is the linearised collision operator.
 
The next step is to attempt a separation of variables into $s_\parallel$ and $\bm{v}/v_\mathrm{T}$, where $\bm{v}$ is made up of parallel and perpendicular components $v_\parallel$ and $v_\perp$, by expressing 
\begin{equation}\label{eq:efunc}
\langle\delta f\rangle = \sum_n A_n(s_\parallel) \psi_n(\bm{v}/v_\mathrm{T}) \textrm{ such that } \frac{\langle C_\mathrm{L}(\psi_n)\rangle}{v_\parallel} = \frac{\psi_n}{\lambda_n},
\end{equation}
\noindent
where $\psi_n$ are eigenfunctions of the operator $\langle C_\mathrm{L}\rangle/v_\parallel$, which depends only on $\bm{v}/v_\mathrm{T}$, and $\lambda_n$ the inverse of their eigenvalue with dimensions of length. Substituting (\ref{eq:efunc}) into (\ref{eq:DKE}) and assuming that the dependence of $\psi_n$ on space through $v_\mathrm{T}$ is negligible (only valid when \color{black}relative temperature perturbations are small globally\color{black}) yields
\begin{equation}
\label{JiHeldFinal}
 \sum_n \Big( \psi_n\frac{\partial A_n}{\partial s_\parallel } +  \cancelto{0}{A_n \frac{\partial \psi_n}{\partial s_\parallel}} + \frac{A_n\psi_n}{\lambda_n} \Big)  = \frac{e E_\parallel}{m_\mathrm{e}v_\parallel}\frac{\partial \langle f^{(0)}\rangle}{\partial v_\parallel}-\frac{\partial \langle f^{(0)}\rangle}{\partial s_\parallel}.
\end{equation}

By similarly decomposing the right-hand side into (orthogonal) eigenfunction contributions, a set of independent first-order ODE's for $A_n$ is formed that can be solved efficiently. Consequently, $\delta f$ can be reconstructed and the nonlocal correction to the heat flux computed through an integral \textcolor{black}{in $v_\parallel$} (hence the nomenclature Eigenvector Integral Closure or EIC).

The advantage of this approach is that it circumvents the need to solve in velocity-space at every timestep (as a VFP code must). The main challenge is identifying a discrete description of the eigenfunctions $\psi_n$ \textcolor{black}{that converges rapidly for use in a numerical scheme.}  In practice, this is done by using an orthonormal polynomial moment-basis to express $\psi_n$ as a vector and the operator $C_\mathrm{L}/v_\parallel$ as a matrix. Ji \emph{et al.} \cite{JiHeld} proposed a Legendre-Laguerre (LL) basis in pitch angle and total speed. This converges rapidly in the hydrodynamic limit but slowly in the collisionless limit. As an alternative, Omotani \emph{et al.} \cite{Hermite} proposed a  Hermite-Laguerre (HL) basis, decoupling parallel and perpendicular velocity components, which  allows for easier implementation of sheath boundary conditions.

\subsection{Non-Fourier Landau-Fluid}

While there are a lot of computational benefits to the EIC model over a full VFP code, a large number of eigenfunctions (at least 120 according to Omotani \emph{et al.}\cite{Hermite}) are needed for convergence. The NFLF model \cite{NFLFTheory,*NFLF2014,NFLFImplementation} provides a cheaper approach, potentially solving as few as three second-order ODE's, but without a direct link to the VFP equation.

The popular Landau-fluid approach initially proposed by Hammett and Perkins \cite{Hammett,H2,H3} provides a closure for the nonlocal heat flux $\bm{\tilde{Q}}$ in Fourier space. This recovers the correct damping rate of a sinusoidal temperature perturbation in both the collisional and collisionless limits (where the wavelength is much longer/shorter than the thermal mfp)\color{black}. However, the Fourier transforms involved are inconvenient for complex SOL geometries and large temperature and density variations.

The innovation by Dimits, Joseph and Umansky \cite{NFLFTheory,*NFLF2014} was to enable direct calculation of the nonlocal parallel heat flux in configuration space by approximating the closure as a sum of Lorentzians
\begin{equation}
\label{eq:Lors}
 \bm{\tilde{Q}} \approx \frac{\bm{\tilde{Q}}^\mathrm{(B)}}{1+a|k|\lambda_\mathrm{ei}^{\textrm{(B)}}} \approx \sum_{j=1}^N \frac{\alpha_j\bm{\tilde{Q}}^\mathrm{(B)}}{\beta_j^2+\big(ak\lambda_\mathrm{ei}^{\textrm{(B)}}\big)^2} \equiv\sum_{j=1}^N \bm{\tilde{q}}_j,
 \end{equation}
where $\bm{\tilde{Q}}^\mathrm{(B)}$ is the (parallel) Braginskii heat flow in reciprocal space, $a$ parametrises the behaviour in the collisionless limit and is determined analytically, $k$ is the wavenumber \textcolor{black}{of the Fourier mode}, $N$ is the number of Lorentzians chosen by the user for the fit and $\alpha_j, \beta_j$ are fit parameters.

Equating the contribution from each Lorentzian to a dummy contribution $\bm{q}_j$, rearranging and taking the inverse Fourier transform gives a set of $N$ second-order ODE's for each spatial direction of interest that can be used to recover the nonlocal heat flow: 
\begin{equation}
\label{eq:NFLF}
\Big(\beta_j^2+\big(ak\lambda_\mathrm{ei}^{\textrm{(B)}}\big)^2\Big)\bm{\tilde{q}}_j \rightarrow \left(\beta_j^2-a^2\lambda_\mathrm{ei}^{\textrm{(B)}2}\nabla^2\right)\bm{q}_j = \alpha_j\bm{Q}^\mathrm{(B)}.
 \end{equation}
\textcolor{black}{This approach also conveniently avoids the issue of defining the mean free path in reciprocal space.}

\subsection{Multigroup Diffusion (SNB)}
The final model being investigated is the multigroup diffusion or `SNB' model named after the original authors Schurtz, Nicola\"i and Busquet \cite{SNB}. It is widely used in inertial fusion codes such as Lawrence Livermore National Laboratory's HYDRA \cite{HYDRA}, CELIA laboratory's CHIC \cite{CHIC} \textcolor{black}{and the University of Rochester Laboratory for Laser Energetics' (LLE) DRACO\cite{Cao}}; and it is applicable in multiple spatial dimensions.

The SNB model is best explained starting from the diffusion approximation of the kinetic equations (see equations (\ref{eq:f0_orig}) and (\ref{eq:f1_orig}) above), \textcolor{black}{ along with neglecting time-derivatives for similar reasons as the EIC model}. The isotropic part of the distribution function $f_0$ is then \color{black} split into a Max\-well-Boltzmann distribution $f_0^{\mathrm{(mb)}}=n_\mathrm{e}\exp(-v^2/2v_\mathrm{T}^2)/(2\uppi v_\mathrm{T})^{3/2}$ and a deviation $\delta f_0=f_0-f_0^{\mathrm{(mb)}}$. The anisotropic part $\bm{f}_1$ is similarly split, but the `Max\-wellian' part $\bm{f}_1^{\mathrm{(mb)}}$, obtained from substituting $f_0^{\mathrm{(mb)}}$ into equation (\ref{eq:f1_orig}), is replaced by an alternative, $\bm{g}_1^{\mathrm{(mb)}}$:
\begin{equation}
\begin{split}
\bm{f}_1^{\mathrm{(mb)}} &= -\lambda_\mathrm{ei}^*  \Big(\frac{m_\mathrm{e} v^2}{2 k_\mathrm{B} T_\mathrm{e}}-4\Big) f_0^{\mathrm{(mb)}} \frac{\nabla T_\mathrm{e}}{T_\mathrm{e}} , \\
\rightarrow \bm{g}_1^{\mathrm{(mb)}} &=  -\lambda_\mathrm{ei}^*   f_0^{\mathrm{(mb)}} \frac{\nabla T_\mathrm{e}}{T_\mathrm{e}}.
\end{split} 
\end{equation}
This modification achieves positive-definiteness without affecting the integral used to calculate the heat flow, and is argued to be compensated by other approximations of the model \cite{SNB}.  Here we have defined  $\lambda_\mathrm{ei}^* = \xi\lambda_\mathrm{ei} = \xi v/\nu_\mathrm{ei}$. Note that the factor of 3 difference from the original paper in $\bm{f}_1^{\mathrm{(mb)}}$ is simply due to the use of spherical harmonics by Schurtz \emph{et al.} while we use a Cartesian tensor expansion.

\textcolor{black}{Electric field terms in equation (\ref{eq:f0_orig}) are neglected and instead incorporated phenomenologically  by defining a limited mfp:
\begin{equation}
\frac{1}{\lambda_\mathrm{ei}^{(E)}} = \frac{1}{\lambda_\mathrm{ei}^*} + \left\vert \frac{e\bm{E}}{\sfrac{1}{2}m_\mathrm{e}v^2}\right\vert.
\end{equation}
}where the local form for $\bm{E}=k_\mathrm{B}T_\mathrm{e}(\nabla \log n_\mathrm{e} + \gamma\nabla \log T_\mathrm{e})$ is used, with $\gamma=1+3(Z+0.477)/2(Z+2.15)$. Substituting $\bm{f}_1 = \bm{g}_1^{\mathrm{(mb)}} + \lambda_\mathrm{ei}^{(E)}\nabla\delta f_0$ into equation the stationary form of (\ref{eq:f0_orig}) obtains
\begin{equation}\label{eq:SNB}
\frac{C_{\mathrm{ee0}}\left[\delta f_0\right]}{v} +\nabla\cdot \frac{\lambda_\mathrm{ei}^{(E)}}{3}\nabla \delta f_0 = \frac{\nabla\cdot \bm{g}_1^{\mathrm{(mb)}} }{3}.
\end{equation}
This can be solved to compute the deviation from the local heat flow
\begin{equation}
\delta \bm{Q} = \frac{2\uppi m_\mathrm{e}}{3}\int_{\mathrlap{0}}^{\mathrlap{\infty}}\delta \bm{f}_1 v^5\;\mathrm{d}v = -\frac{2\uppi m_\mathrm{e}}{3}\int_{\mathrlap{0}}^{\mathrlap{\infty}} v^5\lambda_\mathrm{ei}^{(E)} \nabla \delta f_0\; \mathrm{d}v.
\end{equation}

The main computational advantage of this approach is through the use of efficient model collision operators that are local in velocity-space. This allows for a more effective discretisation into velocity/energy groups, and removes the need to store the entire distribution function in memory. The original authors suggested using a velocity-dependent Krook (BGK) operator due to its simplicity, but Del Sorbo \emph{et al.} \cite{Dario} have also employed a more realistic operator suggested by Albritton, Williams, Bernstein and Swartz (AWBS) \cite{AWBS}:
\begin{align}
C_{\mathrm{ee0}}^\textrm{BGK}[\, \cdot \, ] &= -r\frac{\nu_\mathrm{ei}}{Z}\times \cdot \, , &C_{\mathrm{ee0}}^\textrm{AWBS}[\, \cdot \, ] &= \frac{\nu_\mathrm{ei}}{Z}v\frac{\partial}{\partial v}[\, \cdot\, ],
\end{align}
where \textcolor{black}{we have introduced the dimensionless number $r$ to account for variation in SNB model implementations/description across publications: } The original authors \cite{SNB} halved the geometrically averaged mfp$\lambda_\mathrm{e} = \sqrt{Z\lambda_\mathrm{ei}^*\lambda_\mathrm{ei}}$ (see equation (23) of Schurtz \emph{et al.}\cite{SNB} and also section III C of Del Sorbo \emph{et al.} \cite{Dario}), which is equivalent to setting $r=4$ \textcolor{black}{(except for the treatment of electric field)}. However, in a later section of the original paper\cite{SNB} (III F) as well as section II of a later publication \cite{SNBmag} this technicality is not restated when demonstrating a link to the kinetic equations, from which a value of $r=1$ could be interpreted.

\textcolor{black}{Furthermore, our attempts to replicate previous comparisons between SNB and VFP \cite{Marocchino} suggest that Maro\-{}cchino \emph{et al.}  used $r=16$. Using} this value for $r$ in the SNB model along with neglecting corrections to angular scattering from electron-electron collisions (i.e. $\xi$ is set to one) happens to give good agreement with VFP codes for $Z=1$.  However, this agreement is observed to get progressively worse as $Z$ increases. In this paper we show that using the BGK collision operator with a different value ($r=2$) and $\xi=(Z+0.24)/(Z+4.2)$ gives very good agreement of SNB with VFP across a wide range of problems (and values of $Z$). 

\color{black} Note that, despite the differential form of the AWBS operator, its use does not actually require a significant increase in computational time unless an attempt to parallelise over energy groups is being made. This is because the velocity-space first-order differential equation is simply closed from above with the boundary condition $\delta f_0(v=\infty) = 0$. In a finite difference scheme this could simply be implemented by enforcing the highest energy group to be zero, and solving for the next highest group first. (Bear in mind that discretisation in velocity-space need not be uniform.)  However, we identify other issues with the AWBS operator in section \ref{sec:hydro} that limit its usefulness and the SNB model using this operator is therefore not explored further. \color{black}

\section{Decay of a Small-Amplitude, Sinusoidal Temperature Perturbation}\label{sec:linear}

First we consider the damping of a small-ampli\-{}tude temperature \textcolor{black}{perturbation} $T_e = T_0 + \delta T\cos(kx)$ (often referred to as the Epperlein-Short \cite{EppShort} test) with a constant uniform background density and ionisation.  Due to nonlocal effects as the wavelength is reduced, the dimensionless thermal conductivity $\kappa$ decreases from that predicted in the local limit, $\kappa^\mathrm{(B)}$. \textcolor{black}{In this section we first detail the metho\-{}dology used in setting up simulations of this problem and assessing the respective conductivity reductions before briefly commenting on the agreement between the EIC model and VFP results. Analysis of the long-wavelength limit will then be presented in section \ref{sec:hydro} so as to motivate a suitable choice for the SNB model parameter $r$. Finally, a new fit function for the conductivity reduction as a function of $k\lambda_\mathrm{ei}^\mathrm{(B)}$ is derived in section \ref{sec:cless} by connecting the collisional and collisionless regimes, and is used to calculate fit coefficients for the NFLF model.}

A \textcolor{black}{sinusoidal perturbation} with a relative amplitude of $\delta T/T_0 = 10^{-3}$ was initialised for the KIPP simulations. We used a uniform spatial grid of 127 cells over a half-wavelength with a non-uniform Cartesian velocity grid extending to $v_\mathrm{max} = 7v_\mathrm{T}$ (with parameters $mmax = 256, EPS = 1.01$ as defined by Chankin \emph{et al.}\cite{KIPPDev}). \textcolor{black}{The two methods employed by Marocchino \emph{et al.}\cite{Marocchino} were used to calculate the conductivity reduction $\kappa/\kappa^\mathrm{(B)}$: (1) directly from the peak heat flow divided by the predicted local heat flow ($\kappa/\kappa^\mathrm{(B)}=\tilde{Q}/\tilde{Q}^\mathrm{(B)}$) and (2) inferred from the exponential decay rate $\rho$ of the temperature perturbation ($\kappa/\kappa^\mathrm{(B)}=\rho/\rho^\mathrm{(B)}$, where $\rho^\mathrm{(B)}=2\kappa^\mathrm{(B)}k^2v_\mathrm{T}\lambda_\mathrm{ei}^\mathrm{(B)}/3$).}

 The thermal conductivity obtained by both these methods fluctuated in time initially (due to initialising as a Maxwellian) and was tracked until both methods approached constant values. Due to incomplete convergence in timestep these values were slightly different and an average was then taken between the two final conductivity reductions. In order to improve accuracy without using unnecessary amounts of computational time due to a tiny timestep (KIPP is only first-order accurate in time), extrapolation to zero timestep was performed \color{black}by fitting a straight line of conductivity reduction against timestep.  Such complications were unnecessary when using the inherently stationary models: instead of evolving in time, it was possible to calculate heat flow (and thus conductivity reduction) from a single profile with a \textcolor{black}{lower relative amplitude of $10^{-5}$} for each wavelength.

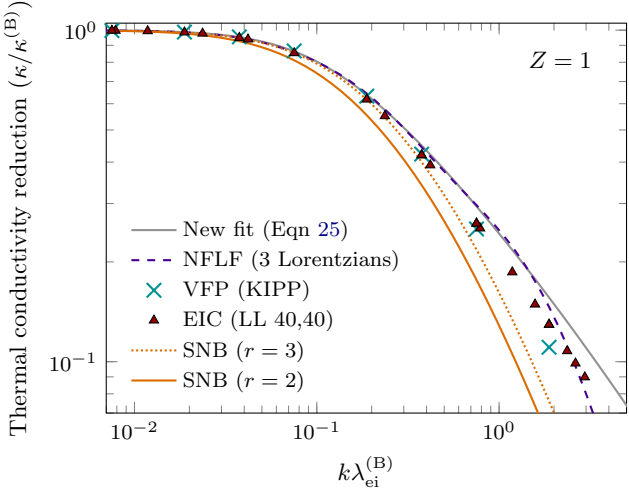
\begin{figure}[t]
\begin{tikzpicture}
\begin{loglogaxis}[
xlabel = {$k\lambda_\mathrm{ei}^\textrm{(B)}$},xmin=0.007,xmax=5.,
ylabel = {Thermal conductivity reduction ($\kappa/\kappa^\mathrm{(B)}$)},ymin=0.07,ymax=1.05,
legend pos = south west,
]
\addplot[themeGray] table {data/linearised_z1_analytical}; \addlegendentry{New fit (Eqn \ref{eq:c1})}
\addplot[nflf] table {data/linearised_z1_nflf3}; \addlegendentry{NFLF (3 Lorentzians)}
\addplot[vfp,only marks,mark=x,mark size=1ex] table {data/linearised_z1_kipp}; \addlegendentry{VFP (KIPP)}
\addplot[thin,black,fill=themeRed,only marks,mark=triangle*] table {data/linearised_z1_legendre4040}; \addlegendentry{EIC (LL 40,40)}
\addplot[snb] table {data/linearised_z1_snbr3}; \addlegendentry{SNB ($r=3$)}
\addplot[snb,solid] table {data/linearised_z1_snbr2}; \addlegendentry{SNB  ($r=2$)}
\node at (rel axis cs:.95,.95) [anchor=north east] {$Z=1$};
\end{loglogaxis}
\end{tikzpicture}
\caption{\label{linear} Reduction of thermal conductivity due to nonlocality over a range of collisionalities for a small-amplitude temperature sinusoid with $Z=1$. The fit function given in equation (\ref{eq:c1}) is depicted in addition to results using the nonlocal models and VFP codes.}
\end{figure} 
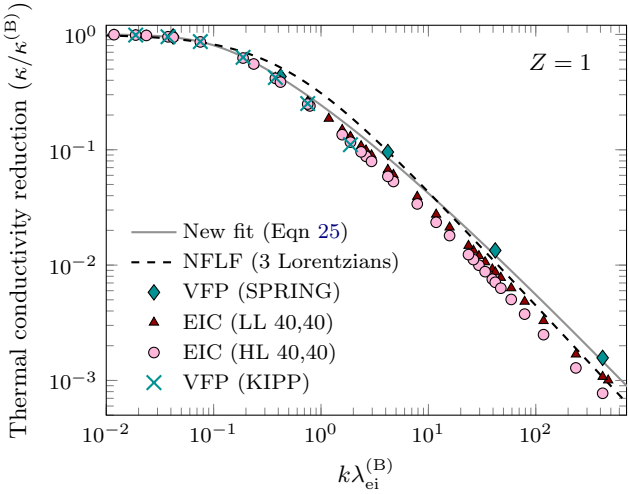
\begin{figure}[t]
\begin{tikzpicture}
\begin{loglogaxis}[
xlabel = {$k\lambda_\mathrm{ei}^\textrm{(B)}$},xmin=0.01,xmax=700.,
ylabel = {Thermal conductivity reduction ($\kappa/\kappa^\mathrm{(B)}$)},ymin=0.0005,ymax=1.2,
legend pos = south west,
]
\addplot[themeGray] table {data/linearised_z1_analytical}; \addlegendentry{New fit (Eqn \ref{eq:c1})}
\addplot[local] table {data/linearised_z1_hp}; \addlegendentry{NFLF (3 Lorentzians)}
\addplot[thin,black,fill=themeGreen,only marks,mark=diamond*,mark size=.75ex] table {data/linearised_z1_spring}; \addlegendentry{VFP (SPRING)}
\addplot[thin,black,fill=themeRed,only marks,mark=triangle*] table {data/linearised_z1_legendre4040}; \addlegendentry{EIC (LL 40,40)}
\addplot[thin,black,fill=themePink!50!white,only marks,mark=*] table {data/linearised_z1_hermite4040}; \addlegendentry{EIC (HL 40,40)}
\addplot[vfp,only marks,mark=x,mark size=1ex] table {data/linearised_z1_kipp}; \addlegendentry{VFP (KIPP)}
\node at (rel axis cs:.95,.95) [anchor=north east] {$Z=1$};
\end{loglogaxis}
\end{tikzpicture}%
\caption{\label{linearhigh} Reduction of thermal conductivity due to nonlocality  extending to lower collisionality for a small-amplitude temperature sinusoid with $Z=1$. SPRING data is reproduced with permission from \citen{Bychenkov,*Bychenkov95}. Copyrighted by the American Physical Society.}
\end{figure} 
\begin{figure}[tph]
\begin{tikzpicture}
\begin{loglogaxis}[
xlabel = {$k\lambda_\mathrm{ei}^\textrm{(B)}$},xmin=0.007,xmax=5.,
ylabel = {Thermal conductivity reduction ($\kappa/\kappa^\mathrm{(B)}$)},ymin=0.01,ymax=1.05,
legend pos = south west,
]
\addplot[themeGray] table {data/linearised_z8_analytical}; \addlegendentry{New fit (Eqn \ref{eq:c1})}
\addplot[vfp,only marks,mark=x,mark size=1ex] table {data/linearised_z8_impact}; \addlegendentry{VFP (IMPACT)}
\addplot[thin,black,fill=themeGreen,only marks,mark=diamond*,mark size=.75ex] table {data/linearised_z8_spring}; \addlegendentry{VFP (SPRING)}
\addplot[thin,black,fill=themeRed,only marks,mark=triangle*] table {data/linearised_z8_legendre5050}; \addlegendentry{EIC (LL 40,40)}
\addplot[snb,solid] table {data/linearised_z8_snbr2}; \addlegendentry{SNB  ($r=2$)}
\node at (rel axis cs:.95,.95) [anchor=north east] {$Z=8$};
\end{loglogaxis}
\end{tikzpicture}%
\caption{\label{linearz8} Reduction of thermal conductivity due to nonlocality over a range of collisionalities for a temperature sinusoid with $Z=8$. SPRING data is reproduced with permission from \citen{Bychenkov,*Bychenkov95}. Copyrighted by the American Physical Society.}
\end{figure}
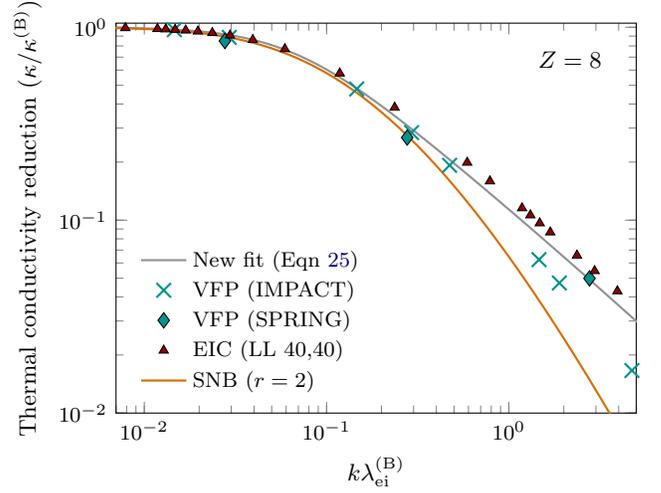 

 \textcolor{black}{Results obtained for thermal conductivity reduction $\kappa/\kappa^\mathrm{(B)}$ as a function of nonlocality parameter $k\lambda_\mathrm{ei}^\mathrm{(B)}$ are shown in Figs. \ref{linear} and \ref{linearhigh} for an ionisation of $Z=1$. The choice of two separate figures for the case of $Z=1$ is to allow for clear identification of features at both high and low collisionality and to avoid an excessive number of comparisons on a single figure.} Kinetic results from the linearised VFP code SPRING calculated by Epperlein\cite{Epperlein94} and provided numerically by Bychenkov \emph{et al.}\cite{Bychenkov,*Bychenkov95} are also shown in Fig. \ref{linearhigh} for comparison against the nonlocal models.

Both the LL\cite{JiHeld} and the HL\cite{Hermite} bases for the EIC model were investigated using 40,40 moments to achieve convergence to within $1\%$ for $k\lambda_\mathrm{ei}^{\textrm{(B)}} < 1$. Figure \ref{linear} shows that both bases agree well with KIPP (within 5 and 10\% respectively) in this limit. For higher $k\lambda_\mathrm{ei}^\mathrm{(B)}$ \textcolor{black}{(see Fig. \ref{linearhigh})}, the SPRING VFP results begin to deviate from both the EIC and KIPP results for a number of reasons: Firstly, the onset of electron inertia effects at high $k\lambda_\mathrm{ei}^\textrm{(B)}$, ignored by the EIC model, introduces a phase shift between the heat flow and temperature perturbation \textcolor{black}{in frequency space (i.e. the perturbation goes from being critically damped to possessing an oscillatory component)}  making evaluation of the decay rate difficult with KIPP (the linearised formulation of the SPRING code makes this easier, and likely more accurate; note that it is the modulus of the complex thermal conductivity that has been provided in this case).

Additionally, \textcolor{black}{while the HL basis only requires two Laguerre modes in the collisionless limit due to the parallel-perpendicular decoupling, we found that even 160 HL modes were insufficent to achieve convergence to within 10\% for $k\lambda_\textrm{ei}^{(\mathrm{B})}>2$. The LL basis, however, manages to converge to within  1\% for $k\lambda_\textrm{ei}^{(\mathrm{B})}<50$ using 20,20 modes.  The collisionless limit predicted by Chang and Callen \cite{ChangCallen} is approached as the total number of LL modes is increased} (see below and also Fig. 2 and 3 of Ji \emph{et al.}\cite{JiHeld} whose results we have successfully replicated), this is about a factor of 1.8 less than the true collisionless heat flow predicted analytically and by the SPRING code (see section \ref{sec:cless}). 

\begin{table*}[t] 
 \vspace{-1em}
  \caption{\label{tab:bZ} Values for the parameter $b$, as appearing in equation (\ref{eq:bZ}), characterising lowest-order deviation from hydrodynamic limit for various values of $Z$ obtained with the EIC model. At least 4,40 moments were used in the LL basis.}
 \begin{ruledtabular}
 \begin{tabular}{c|ccc   c    c    c     c    c   c  c   c  c   c}
 $Z$ &1     &2    &3    &4   &6   &8   &10  &12 &14 &20&30&500&$\infty$\\ 
 $b$ & 43.5 &73.6 &96.0 &113 &139 &157 &170 &180&189&206&222&261&264  \\
 \end{tabular}
 \end{ruledtabular}
\end{table*}

Results for an ionisation of $Z=8$ are shown in Fig. \ref{linearz8}. \textcolor{black}{Here 50,50 moments in the LL basis were required to achieve convergence at high $k\lambda_\textrm{ei}^{\textrm{(B)}}$ with the EIC. The diffusion approximation made by IMPACT is shown to  break} down around $k\lambda_\textrm{ei}^{\textrm{(B)}}\approx0.3$. Note that the thermal conductivity reduction \textcolor{black}{at which the SNB begins to deviate from kinetic results is about the same ($\kappa/\kappa^\mathrm{(B)}\approx0.2$) for both $Z=1$ and 8;} the lower nonlocality parameter $k\lambda_\textrm{ei}^{\textrm{(B)}}$ at which this occurs is due to the reduced importance of electron-electron collisions at higher ionisations.

\subsection{Hydrodynamic Limit \texorpdfstring{($k\lambda_\mathrm{ei}^\mathrm{(B)} \ll 1$)}{} }\label{sec:hydro}

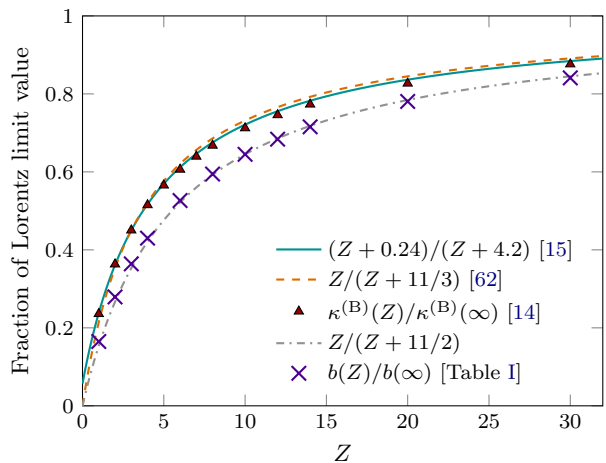
\begin{figure}[b]
\begin{tikzpicture}
\begin{axis}[
xlabel = {$Z$},xmin=0,xmax=32,
ylabel = {Fraction of Lorentz limit value},ymin=0.,ymax=1.,
legend pos = south east
]
\addplot[themeGreen] table {data/z_xi_EppShort}; \addlegendentry{$(Z+0.24)/(Z+4.2)$ \citen{EppShort}}
\addplot[themeOrange,dashed] table {data/z_xi_Sanmartin};\addlegendentry{$Z/(Z+11/3)$ \citen{Sanmartin}}
\addplot[thin,black,fill=themeRed,only marks,mark=triangle*] table {data/z_xi_Actual_EppHaines};\addlegendentry{$\kappa^\mathrm{(B)}(Z)/\kappa^\mathrm{(B)}(\infty)$ \citen{EppHaines}}
\addplot[themeGray,dashdotted] table {data/z_b_newfit};\addlegendentry{$Z/(Z+11/2)$}
\addplot[themePurple,only marks,mark=x,mark size=1ex] table {data/z_b_Actual_EIC};\addlegendentry{$b(Z)/b(\infty)$ [Table \ref{tab:bZ}]}
\end{axis}
\end{tikzpicture}%
\caption{\label{fig:bZ} Comparison of the $Z$-dependence of the local thermal conductivity $\kappa^\mathrm{(B)}$ and the parameter $b$ in equation (\ref{eq:bZ}), which characterizes the nonlocal deviation from the local limit.}
\end{figure} 

\color{black} Bychenkov \emph{et al.} \cite{Bychenkov,*Bychenkov95} have shown that for long wavelength perturbations (i.e. in the hydrodynamic limit)
\begin{equation} \label{eq:bZ}
\bm{\tilde{Q}} \sim \bm{\tilde{Q}}^\mathrm{(B)}\!\big(1-bZk^2\lambda_\mathrm{ei}^{\textrm{(B)}2}\big)
\end{equation}

\noindent to second-order in $k\lambda_\mathrm{ei}^{\textrm{(B)}}$, where $b\approx264$ in the Lorentz limit ($Z=\infty$). As the assumptions of the EIC model are valid in this linear and collisional limit (except for neglection of electron inertia which would only introduce a \textcolor{black}{time-dependent} component into the heat flow), and convergence of the LL basis is relatively rapid (only 2 Legendre modes are theoretically needed) we have used it to calculate the value of $b$ for various $Z$ \textcolor{black}{(while the KIPP prediction for $Z=1$ was within 4\% of the EIC, this was considered less accurate due to insufficient \textcolor{black}{extension/resolution of} the velocity grid)}. This was done by fitting a straight line on a graph of heat flow against $Zk^2\lambda_\mathrm{ei}^{\textrm{(B)}2}$ for $k\lambda_\mathrm{ei}^{\textrm{(B)}} <10^{-3}/\sqrt{Z}$ (the lower range of  $k\lambda_\mathrm{ei}^{\textrm{(B)}}$ extended below $2\times 10^{-4}/\sqrt{Z}$ and there were typically at least six points on the graph).

Results using the EIC model are summarised in Table \ref{tab:bZ} and Fig. \ref{fig:bZ}, which also includes numerical results \cite{EppHaines} and rational approximations \cite{EppShort,Sanmartin} for the low-Z conductivity correction \textcolor{black}{$\kappa^\mathrm{(B)}(Z)/\kappa^\mathrm{(B)}(\infty)$}. We find that the approximation $b(Z)/b(\infty)=Z/(Z+11/2)$  fits numerical results to within 7\%, whereas simply using $\xi$ overestimates $b$ by as much as 43\% at $Z=1$. \color{black} However, the implications of this for the validity of using $\xi$ in IMPACT and the SNB model are not as serious as they seem because $b$ only quantifies the initial deviation from the local limit, and the total heat flux is not very sensitive to marginal errors in $b$ in the hydrodynamic limit.  \color{black}

Table \ref{tab:SNB} outlines the values of $b$ predicted by the SNB model depending on the model collision operator and choice of source term. This has been derived in the low-amplitude and continuum-velocity limit as detailed in Appendix A.  Using the AWBS operator and the kinetic source term $\nabla \cdot \bm{f}_1^{\mathrm{(mb)}}$ on the right-hand side of equation (16) gives \emph{a priori} the closest value of $b=316.9\xi$ (top right) to within 20\% of  that predicted analytically in the Lorentz limit (Table 1)

 The ability of the AWBS collision operator to predict the deviation in the hydrodynamic limit fairly accurately might suggest that it provides an improvement to the original SNB model, however we find that coupling it with the original source term leads to negative values of the thermal conductivities at $k\lambda_\mathrm{ei}^{\textrm{(B)}}> 0.124/\sqrt{\xi Z}$ due to it not being positive-definite (see Appendix B). This should never occur in the linearised problem considered here (i.e. decay of a small-amplitude temperature perturbation) as it would result in instabilities at these wavelengths. However, this issue does not necessarily imply that the AWBS operator is an inappropriate choice for \textit{other} nonlocal models. For example, the M1 model presented by Del Sorbo \emph{et al.}\cite{Dario,Dariomag} does not appear to exhibit this issue of positive-definitiveness; we leave a detailed analysis of this model for future work.

Setting $r=2$ exactly in the original implementation of the SNB model (BGK collision operator with the source term $\nabla \cdot \bm{g}_1^{\mathrm{(mb)}}$) remarkably gives the same value of $b=316.9\xi$ as with the AWBS operator and the source term $\nabla \cdot \bm{f}_1^{\mathrm{(mb)}}$ (compare bottom left and top right entries of Table \ref{tab:SNB}) and in fact the entire distribution function in this limit (see Appendix A). However, to match the kinetic results for $b$, a value of $r=2.4$ is required in the Lorentz limit and $r=3$ for $Z=1$. We suggest that matching coefficients to such accuracy is not necessary, and that using $r=2$ achieves much better agreement for problems involving large temperature variations (see below). Results using both $r=2$ and $r=3$ for $Z=1$ have been provided in figures to enable the reader to compare. 

\begin{table}[b] 
 \vspace{-1em}
  \caption{\label{tab:SNB} Predictions for $b$ by the SNB model, depending on choice of collision operator (columns) and source term (rows)}
 \begin{ruledtabular}
 \begin{tabular}{c|cc}
 RHS & $C_{\mathrm{ee0}}^\textrm{BGK}$         & $C_{\mathrm{ee0}}^\textrm{AWBS}$ \\
                       \hline
 $\nabla \cdot \bm{f}_1^{(\textrm{mb})}$ & $3169\xi/r$ &$316.9\xi$  \\ 
 $\nabla \cdot \bm{g}_1^{(\textrm{mb})}$ &$633.8\xi/r$   &$63.38\xi$ \\
 \end{tabular}
 \end{ruledtabular}
\end{table}

Faithfulness to kinetic results for $b$ can be guaranteed with the NFLF model by modifying the analytical Fourier closure and constraining the fit coefficients appropriately as described in the next section.

\subsection{Collisionless Limit \texorpdfstring{($k\lambda_\mathrm{ei}^\mathrm{(B)}\gg1$)}{}}\label{sec:cless}

With decreasing wavelength, the heat flow is predicted to slowly approach a constant value. By fitting to the results of both the EIC and SPRING models. \color{black}(we were unable to obtain meaningful KIPP results at low enough collisionalities due to issues mentioned above) \color{black} we find that 
\begin{equation}\label{eq:cless}
\bm{\tilde{Q}} \sim \frac{3}{2}\sqrt{2} \chi_1 \bm{\tilde{Q}}_\mathrm{fs} \left(1-\frac{c_\infty}{ k^{\epsilon}}\right)\frac{\delta T}{T_0},
\end{equation}
where $\bm{\tilde{Q}}_\mathrm{fs}$ is antiparallel to the wave-vector \textcolor{black}{and $\chi_1$, $c_\infty$ and $\epsilon$ are dimensionless fit parameters}, is a reasonable description for the asymptotic behaviour in this limit for low-$Z$ plasmas \color{black} (i.e. a graph of $\tilde{Q}$ against $\log(k)$ resembles a straight line). \textcolor{black}{The form of this fit function was taken from work by Bychenkov \emph{et al.}\cite{Bychenkov,*Bychenkov95}} The extra factor $\sfrac{3}{2}$ compared to the formalism of Hammett and Perkins\cite{Hammett} (which inspired previous implementations of the NFLF model\cite{NFLFImplementation,NFLFTheory,*NFLF2014}) was found to be necessary due to the isotropic definition of the electron temperature used here. \cite{Epperlein94,Bychenkov,*Bychenkov95}

Again, the LL basis was used, this time due to the nonconvergence of the HL basis explained above, however at least 40,40 moments were needed to achieve convergence above $k\lambda_\mathrm{ei}^\mathrm{(B)}\approx1$. As found by the original developers of the model\cite{JiHeld}, the value of $\chi_1=1.2/\sqrt{\uppi}$ agreed with the value predicted by Chang and Callen\cite{ChangCallen}; this is exactly 40\% less than the value predicted by Hammett and Perkins \cite{Hammett} ($\chi_1=2/\sqrt{\uppi}$) because of the quasi-stationary assumption. We have also calculated that an alternative value of $\chi_1=1.225$ can be obtained by numerically solving for zeroes of the response function.

\begin{table}[b] 
 \vspace{-1em}
 \caption{\label{tab:cless} Values for parameters appearing in equations (\ref{eq:cless}) and (\ref{eq:c1}) obtained with EIC model (using at least 40,40 LL moments) and available SPRING data from reference \citen{Bychenkov,*Bychenkov95} (in parentheses), the latter is presumed to be more accurate.}
 \begin{ruledtabular}
 \begin{tabular}{c|ccccc}
$ Z$ &1 &2& 4 & 6 & 8  \\  
 $\epsilon$ &0.32 (0.32) &0.28 & 0.23& 0.22 & 0.20 (0.19) \\
 $c_\infty$ & 0.6 (1.4) & 0.7 &  0.7 & 0.75 & 0.75 (1.5)\\
 $c_1$ & 1.9 (1.5) & 2.2 &2.7 & 3.1 & 3.4 (3.0) \\
 \end{tabular}
 \end{ruledtabular}
 \end{table}

Calculated values of $\epsilon$ and $c_\infty$, as well as the $c_1$ referred to below, are summarised in Table \ref{tab:cless} for $Z=1,2,4,6,8$. Simulations with EIC at higher $Z$ require a prohibitive number of moments for convergence at high $k\lambda_\mathrm{ei}^{\textrm{(B)}}$. Both the index $\epsilon$ and the coefficient $c_\infty$ vary weakly with $Z$ and have similar orders of magnitude to those predicted by Bychenkov \emph{et al}\cite{Bychenkov,Bychenkov95}. The values obtained here should be slightly closer to reality as Bychenkov \emph{et al.} assume complete stationarity (all time derivatives neglected) in their calculations, but there are large uncertainties in our numerical fit (approximately 10\% for the EIC data). The limited numerical results available from the assumingly exact SPRING code\cite{Epperlein94,Bychenkov,*Bychenkov95} infer a value for $\epsilon$ at $Z=1$ within $0.5\%$ of the EIC prediction, but the value for $c_\infty$ (=1.36) is larger by a factor of 2.2.

Due to the combination of stationarity and diffusion approximations, the SNB model without the phenomenological mfp limitation to include electric fields predicts the collisionless heat flow to decrease as $\mathord{\sim}1/k$ to zero as the wavelength decreases\cite{SNB} \textcolor{black}{(the thermal conductivity correspondingly decreases as $1/k^2$)}. Incorporating the mfp limitation resolves the issue of \textcolor{black}{insufficiently damping temperature perturbations of \emph{finite} amplitude (such that $k\lambda_\mathrm{ei}\delta T\gtrapprox 1$). This improves} numerical stability, but introduces an amplitude-dependence of $\chi_1$ that is not observed in VFP simulations.

 While the NFLF will also always predict a $\mathord{\sim}1/k$ decay of the heat flow for high enough $k\lambda_\mathrm{ei}^{\textrm{(B)}}$, increasing the number of Lorentzians used to improve the fit can progressively extend the validity into lower collisionality regimes. The fitting function we used interpolates behaviour in both the hydrodynamic and collisionless limits with a similar but slightly more robust method than used by Bychenkov \emph{et al.}\cite{Bychenkov,*Bychenkov95}:
\begin{equation}\label{eq:c1}
\frac{\kappa}{\kappa^\mathrm{(B)}} = \bigg(1+\bigg(\frac{1}{bZk^2\lambda_\mathrm{ei}^{\textrm{(B)}2}}+\frac{\sfrac{3}{2}\sqrt{2}\chi_1/\kappa^\mathrm{(B)}}{k\lambda_\mathrm{ei}^{\textrm{(B)}} (1+c_1/k^\epsilon)}\bigg)^{-1} \bigg)^{-1},
\end{equation}
where $c_1$ differs from $c_\infty$ by optimising the fit for $k\lambda_\mathrm{ei}^{\textrm{(B)}} \leqslant 1$. Using the parameters as defined in Table \ref{tab:cless} for $Z=1$, \textcolor{black}{equation (\ref{eq:c1})} fits the KIPP and SPRING results to within $6$ and $10\%$ respectively for $k\lambda_\mathrm{ei}\leqslant 1$ and up to $26/20\%$ above this; altering the value of $c_1$ to 1.5 reduces the maximum discrepancy with SPRING results to 11\%.

 \begin{table}[b]
 \vspace{-1em}
 \caption{\label{tab:3L} NFLF fit parameters for $N=3,6$ ($Z=1$)}
 \begin{ruledtabular}
 \begin{tabular}{c|cccccc}
 $N$&3 && &&&\\
$ \alpha$ &$2.176\mathord{\times}10^{-3}$ & 0.06316 &   1.6823 &&& \\  
 $\beta$ &0.1020 &0.3513  &2.4554 &&&
     \\
     $N$&6 && &&&\\
$ \alpha$ & $1.575\mathord{\times}10^{-4}$ & 0.01206 &  0.07960 & 0.5086 & 3.5041 &        49.3331 \\ 
 $\beta$ & 0.06195 &  0.17684 &  0.5064 &  1.7432      & 7.0442& 44.4953  \\
 \end{tabular}
 \end{ruledtabular}
\end{table}

This new fit is depicted in Fig. 1 with the simpler fit $1/(1+ak\lambda_\mathrm{ei}^{\textrm{(B)}})$ \textcolor{black}{obtained by Hammett and Perkins \cite{Hammett} previously used in the NFLF model \cite{NFLFTheory,*NFLF2014}, ($a$ can be related to $\chi_1$ by $a=2\kappa^\mathrm{(B)}/3\sqrt{2}\chi_1$)}, which overestimates the thermal conductivity at moderate collisionalities around $k\lambda_\mathrm{ei}^{\textrm{(B)}} \approx 0.5$ by over $25\%$. \color{black}Note that we have observed a recent closure in configuration space (thus convenient for convolution models) suggested by Ji and Held\cite{JiClosure} to closer fit the EIC results with one more fitting parameter (if  the $\alpha$ used by Ji and Held is not considered a free parameter)---tuning of these parameters could probably also achieve an improved fit to the VFP results. We would also like to highlight a recent paper by Joseph and Dimits who have performed detailed analysis of closures for the response function that connects the collisionless and collisional regime\cite{Joseph}.\color{black}

Three Lorentzians (i.e. $N=3$ in equation \ref{eq:Lors}) can approximate our new fit to within $2.5\%$ up to $k\lambda_\mathrm{ei}^{\textrm{(B)}}\approx1.6$; using six Lorentzians allows this to be extended up to $k\lambda_\mathrm{ei}^{\textrm{(B)}}\approx30$. The coefficients used are given in Table \ref{tab:3L}, and were obtained using the variable projection method \cite{varpro}, constrained such that equation (\ref{eq:bZ}) is obeyed to second-order in $k\lambda_\mathrm{ei}^{\textrm{(B)}}$. 

\section{Comparison for large temperature variations}

\subsection{Homogeneous density and ionisation}\label{sec:homo}

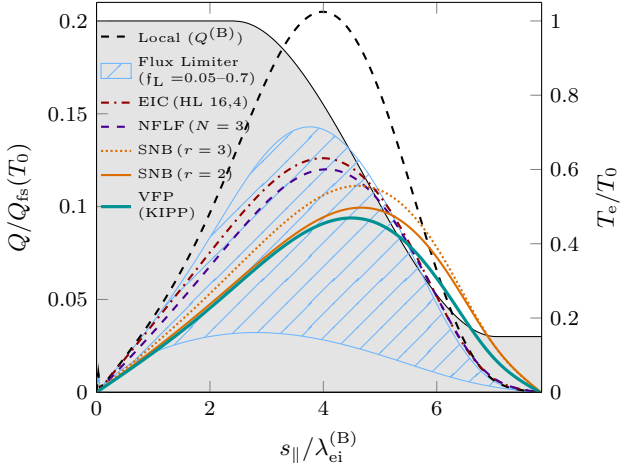
\begin{figure}[t]
\begin{tikzpicture}[trim axis left]
\tikzset{
        hatch distance/.store in=\hatchdistance,
        hatch distance=10pt,
        hatch thickness/.store in=\hatchthickness,
        hatch thickness=0.4pt,
    }

    \makeatletter
    \pgfdeclarepatternformonly[\hatchdistance,\hatchthickness]{flexible hatch}
    {\pgfqpoint{0pt}{0pt}}
    {\pgfqpoint{\hatchdistance}{\hatchdistance}}
    {\pgfpoint{\hatchdistance-1pt}{\hatchdistance-1pt}}%
    {
        \pgfsetcolor{\tikz@pattern@color}
        \pgfsetlinewidth{\hatchthickness}
        \pgfpathmoveto{\pgfqpoint{0pt}{0pt}}
        \pgfpathlineto{\pgfqpoint{\hatchdistance}{\hatchdistance}}
        \pgfusepath{stroke}
    }
\pgfplotsset{
    /pgfplots/layers/my layers/.define layer set=
    {axis background,pre main,my layer,main,axis grid,axis ticks,axis lines,axis tick labels,%
    axis descriptions,axis foreground}
   {/pgfplots/layers/standard},
    set layers=my layers,
    every axis/.append style={scale only axis,width=0.8\linewidth-1cm,height=0.6\linewidth,},
    scaled ticks = false, 
    xmin=0, xmax=7.85,
    legend image code/.code={
\draw[mark repeat=2,mark phase=2]
plot coordinates {
(0cm,0cm)
(0.2cm,0cm)        
(0.4cm,0cm)         
};%
}
}
\begin{axis}[
  axis y line*=right,
  ymin=0, ymax=1.05,
  xlabel={$s_\parallel/\lambda_\mathrm{ei}^\mathrm{(B)}$},
  ylabel={$T_\mathrm{e}/T_0$},
    set layers=my layers,
    clip mode=individual,
]
 \addplot+[thin,black,fill=themeGray!25!white,mark=none,on layer=pre main] table {data/Fig5_xmfp_Teinit} \closedcycle;
\end{axis}

\begin{axis}[
    set layers=my layers,
  axis y line*=left,
  axis x line=none,
  ymin=0, ymax=0.21,
  ylabel={$Q/Q_\mathrm{fs}(T_0)$},
  legend style={font=\tiny,at={(-0.0125,0.965)},anchor=north west},
]
\addplot[local] table {data/Fig5_xmfp_qlocal};
\addlegendentry{Local ($Q^\mathrm{(B)}$)}
\addplot[draw=none,mark=none,name path=low,forget plot] table {data/Fig5_xmfp_qfl_f0.05};
\addplot[draw=none,mark=none,name path=high,forget plot] table {data/Fig5_xmfp_qfl_f0.7};
\addplot[draw,themeBlue,pattern=flexible hatch,pattern color=themeBlue,on layer=my layer,forget plot] fill between[of=high and low];
\addlegendimage{legend image code/.code={%
                                  \draw[thin,themeBlue,pattern=flexible hatch,pattern color=themeBlue] 
                                  (0cm,-0.1cm) rectangle (0.4cm,0.1cm);
}}
\addlegendentry{Flux Limiter\\($\mathfrak{f}_\mathrm{L}=$0.05--0.7)}
\addplot[eic] table {data/Fig5_xmfp_qEICHermite164}; 
\addlegendentry{EIC\,(HL 16,4)}
\addplot[nflf] table {data/Fig5_xmfp_qNFLF3}; 
\addlegendentry{NFLF\,($N=3$)}
\addplot[snb] table {data/Fig5_xmfp_qSNBr3}; 
\addlegendentry{SNB\,($r=3$)}
\addplot[snb,solid] table {data/Fig5_xmfp_qSNBr2}; 
\addlegendentry{SNB\,($r=2$)}
\addplot[vfp,very thick] table {data/Fig5_xmfp_qKIPP}; 
\addlegendentry{VFP\\(KIPP)}
\end{axis}
\end{tikzpicture}%
\caption{\label{heatandtemp} Initial temperature profile (shaded grey) and heat flow as ratio of free-streaming limit $Q_\mathrm{fs} =  n_\mathrm{e} k_\mathrm{B} T_\mathrm{e} v_\mathrm{T}$ (for electrons with energy $k_\mathrm{B}T_0$) after 2.7 collision times. NFLF used 6 Lorentzians, and EIC used 16,4 HL moments.}
\end{figure} 

We now investigate the accuracy of the EIC, NFLF and SNB models in calculating the heat flow in the case where we have a large relative temperature variation. We consider the case of an initial temperature profile consisting of a ramp connecting two large hot and cold regions (`baths'). This has the advantages of allowing simple reflective boundary conditions and not requiring any external heating/cooling mechanisms that would also need to be carefully calibrated between codes. Results and initial conditions are here presented in terms of reference quantities encouraging the translation of the problem to both ICF and MCF relevant situations.

The hot and cold baths had temperatures of $T_0$ and $0.15 T_0$; these were connected by a cubic  ramp given by \color{black}
\begin{equation}
T_\mathrm{e} /T_0 = \begin{dcases}
 1  &  n_\mathrm{c}' \leqslant -75\\
0.575 - \frac{0.85}{300}
n_\mathrm{c}'\bigg(3-\left(\frac{n_\mathrm{c}'}{75}\right)^2\bigg) & n_\mathrm{c}' \in [-75,75] \\
0.15 & n_\mathrm{c}'  \geqslant 75,
\end{dcases}
\end{equation}
where $n_\mathrm{c}' \in [-154,100]$ is the cell number counting from the centre of the temperature ramp. \color{black} Cell size in mfp's was varied between simulations to scan a range of collisionalities. The initial temperature profile is illustrated in Fig. \ref{heatandtemp} \textcolor{black}{for the smallest cell-size used}. For these simulations the electron density, Coulomb logarithm and ionisation ($Z$ = 1) were all kept constant and uniform.

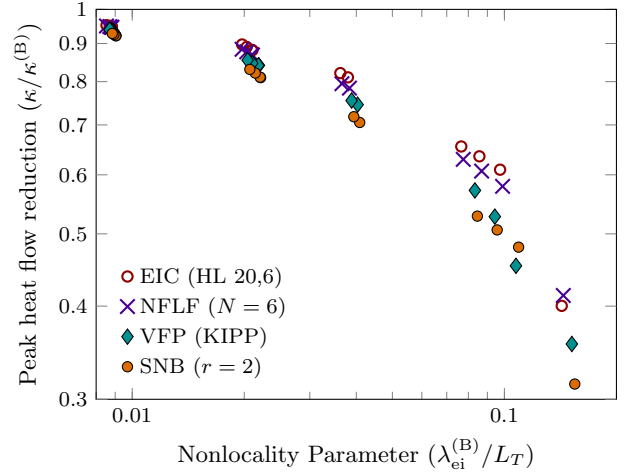
\begin{figure}[t]
\begin{tikzpicture}
\begin{loglogaxis}[
xlabel = {Nonlocality Parameter ($\lambda_\mathrm{ei}^\mathrm{(B)}/L_T$)},xmin=0.008,xmax=0.2,
ylabel = {Peak heat flow reduction ($\kappa/\kappa^\mathrm{(B)}$)},ymin=0.3,ymax=1.,
ytick={1.0},
extra y ticks={0.3,0.4,0.5,0.6,0.7,0.8,0.9},
minor ytick={},
log ticks with fixed point,
yticklabel={\pgfmathfloatparsenumber{\tick}\pgfmathfloatexp{\pgfmathresult}\pgfmathprintnumber{\pgfmathresult}},
legend pos = south west
]
\addplot[themeRed,only marks,mark=o,mark size=.5ex] table {data/Fig6_xmfp_qoverqlocal_EICHermite206}; \addlegendentry{EIC (HL 20,6)}
\addplot[themePurple,only marks,mark=x,mark size=1ex] table {data/Fig6_xmfp_qoverqlocal_NFLF6};\addlegendentry{NFLF ($N=6$)}
\addplot[thin,black,fill=themeGreen,only marks,mark=diamond*,mark size=.75ex] table {data/Fig6_xmfp_qoverqlocal_KIPP};\addlegendentry{VFP (KIPP)}
\addplot[thin,black,fill=themeOrange,only marks,mark=*,mark size=.5ex] table {data/Fig6_xmfp_qoverqlocal_SNBr2};\addlegendentry{SNB ($r=2$)}
\end{loglogaxis}
\end{tikzpicture}%
\caption{\label{allscales} Ratio of peak heat flow to that predicted classically for each snapshot against inverse scalelength $\lambda_\mathrm{ei}/L_T$ (calculated at the location of maximum heat flow  predicted by each model) for the nonlinear temperature ramp using different initial gradients.}
\end{figure} 

KIPP simulations showed an evolution of the heat flow from the local (due to initialising as a Maxwellian) to the nonlocal, with a reduced peak, over an initial transient phase (over which the temperature ramp flattened somewhat). The transient phase was considered over when the ratio of the KIPP heat flow to the expected local heat flow stopped decreasing.  After the transient phase this ratio begins to slowly increase as the thermal conduction flattens the temperature ramp and the ratio of the scalelength to mfp increases (i.e. the thermal transport slowly becomes more local).  We then took the temperature profile from KIPP and used our implementation of the various nonlocal models to calculate the heat flow they would predict given this profile.

Figure \ref{heatandtemp} shows that the EIC and NFLF models agree well with each other (to within $10\%$ almost everywhere for the snapshot shown). However, agreement with KIPP is not nearly as good; the models overestimate the peak heat flux by 30--35\% and do not predict the observed preheat into the cold region. The SNB model is shown to perform much better here, predicting the peak heat flux to within $6\%$ as well as the existence of preheat (although this is overestimated). 

\textcolor{black}{The wide range of heat flow profiles predicted with different flux-limiters between 0.05 and 0.7 are also shown in Fig. \ref{heatandtemp}. These were obtained using the formula $1/Q_\mathrm{fl}=1/Q^\mathrm{(B)}+1/\mathfrak{f}_\mathrm{L}Q_\mathrm{fs}$.  We find that a flux-limiter of $\mathord{\sim}0.25$ best matches the peak kinetic heat flow, but in this case the peak is shifted towards the hot rather than the cold bath (the latter is observed in the KIPP simulation).} Similar results are observed at all temperature ramp scalelengths investigated as illustrated in Fig. \ref{allscales}, which depicts the reduction in the peak heat flow compared to the local Braginskii prediction.

\subsection{Spatially-varying density and ionisation}\label{sec:varion}

While comparisons between the SNB model and VFP codes have previously been performed \cite{SNB,Marocchino}, none have included spatially-inhomogeneous ionisation. \textcolor{black}{As inertial fusion experiments involve steep ionisation and density gradients (for example, at the interface between the helium gas-fill and the ablated gold plasma), it is critical that the SNB model} be tested in such an environment. Variations in ionisation may also be important in the `detached' divertor scenario where a moderate-$Z$ gas is injected in front of the divertor to radiate excess heat; an investigation of this scenario is left as further work. \color{black}For evaluating this, the IMPACT\cite{IMPACT} VFP code was used due to its ability to \textcolor{black}{simulate inhomogeneous ionisation profiles}.\color{black}

\begin{figure}[t]
\begin{tikzpicture}[trim axis left]
\pgfplotsset{
    xmin=1000, xmax=1900,
    scaled x ticks={real:1000.},
    xtick scale label code/.code={},
    every axis/.append style={
    scale only axis,width=0.8\linewidth-1cm,height=0.6\linewidth,
    restrict x to domain=900:1910,label style={font=\small}}
}
\begin{axis}[
  axis y line*=left,
  ymin=0, ymax=3.,
  xlabel={$s$ (mm)},
  ylabel={$T_e$ (keV)},
  axis on top
]
\addplot[themeGray,mark=none] table {data/gdhohlraum_xmic_init_TekeV_interp} [yshift=8pt] node[pos=0.2,anchor=west] {$T_{\textrm{e}}$(20 ns)};
\addplot[themeRed,dashed,mark=none] table {data/gdhohlraum_xmic_5ps_TekeV_interp}[yshift=-12pt] node[pos=0.2,anchor=west] {$T_\mathrm{e}$(20.005 ns)};;
\addplot[fill=themeGray,mark=*,mark size=.25ex,only marks] table {data/gdhohlraum_xmic_init_TekeV_hydra};
\end{axis}

\begin{axis}[
  axis y line*=right,
  axis x line=none,
  ymin=0, ymax=80.,
  ylabel={$Z$ $\vert$ $n_\mathrm{e}$ $(10^{20}$ cm\textsuperscript{$-3$})},
]
\addplot[themeGreen,dashdotted,mark=none] table {data/gdhohlraum_xmic_Z_interp}[yshift=8pt] node[pos=0.85] {$Z$};
\addplot[thin,black,fill=themeGreen,mark=triangle*,only marks,mark size=.5ex] table {data/gdhohlraum_xmic_Z_hydra};
\addplot[themePurple,mark=none] table {data/gdhohlraum_xmic_ne1e20cm3_interp}[yshift=8pt] node[pos=0.85] {$n_{\textrm{e}}$};
\addplot[thin,black,mark=text,text mark=$\mathbf{\vert}$,text mark as node=true,text mark style={black,scale=0.5},only marks] table {data/gdhohlraum_xmic_ne1e20cm3_hydra};
\end{axis}
\end{tikzpicture}%
\caption{\label{fig:initVar} Temperature, density and ionisation profiles after 20 ns simulated laser heating with HYDRA (marks). Curves show interpolated profiles used to initialise IMPACT simulations, as well as the temperature profile after a further 5 ps.}
\end{figure}
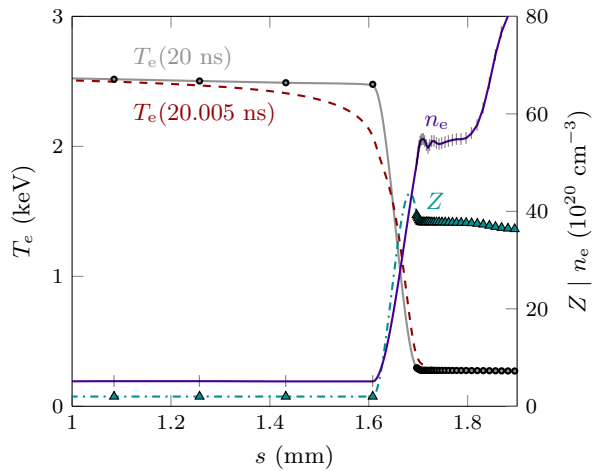 

We performed a HYDRA simulation in 1D spherical geometry of a laser-heated gadolinium hohlraum containing a typical helium gas-fill. A leak source was implemented with an area equal to the laser entrance hole to reproduce the energy balance. Electron temperature ($T_\mathrm{e}$), density ($n_\mathrm{e}$) and ionisation ($Z$) profiles (shown in Fig. \ref{fig:initVar}) at 20 nanoseconds were extracted and used as the initial conditions (along with the assumption that the electron distribution function is initially Maxwellian everywhere) for the IMPACT simulation (which was performed instead in planar geometry). At this point very steep gradients in all three variables were set up with a change from $T_\mathrm{e} = 2.5\ \mathrm{keV}, n_\mathrm{e} = 5\mathord{\times}10^{20}\ \mathrm{cm}^{-3}, Z=2$ to $T_\mathrm{e} = 0.3\ \mathrm{keV}, n_\mathrm{e} = 5\mathord{\times}10^{21}\ \mathrm{cm}^{-3}, Z=39$ across approximately 100~$\upmu$m at the helium-gadolinium interface.
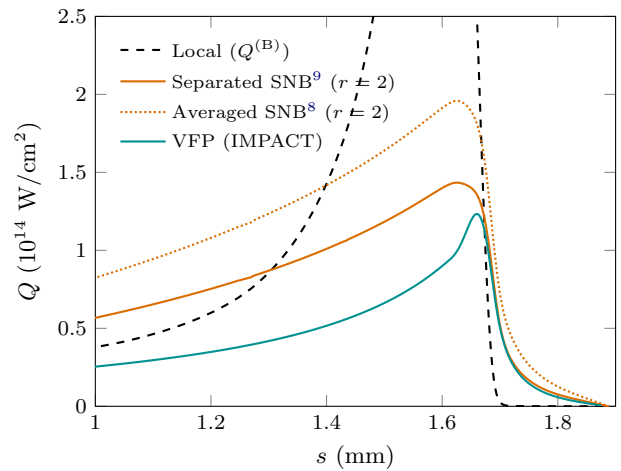
\begin{figure}[t]
\begin{tikzpicture}
\begin{axis}[
    xmin=1000, xmax=1900,
    scaled y ticks={real:1e14},
    ytick scale label code/.code={},
    ymin=0,ymax=2.5e14,
    scaled x ticks={real:1000.},
    xtick scale label code/.code={},
    restrict x to domain=900:1910,label style={font=\small},
    xlabel={$s$ (mm)},
    ylabel={$Q$ (10\textsuperscript{14} W/cm\textsuperscript{2})},
    legend pos = north west,
    legend style={font=\scriptsize},
]
\addplot[local] table {data/gdhohlraum_xmic_5ps_LocalWcm2};
\addlegendentry{Local ($Q^\mathrm{(B)}$)}
\addplot[snb,solid] table {data/gdhohlraum_xmic_5ps_separatedsnbWcm2} ;
\addlegendentry{Separated SNB\cite{SNBmag} ($r=2$)} 
\addplot[snb] table {data/gdhohlraum_xmic_5ps_averagedsnbWcm2};
\addlegendentry{Averaged SNB\cite{SNB} ($r=2$)} 
\addplot[vfp] table {data/gdhohlraum_xmic_5ps_IMPACTWcm2};
\addlegendentry{VFP (IMPACT)}
\end{axis}
\end{tikzpicture}%
\caption{\label{fig:Zvar} Comparison of heat flow predictions with the SNB model using geometrically averaged or separated mfp's based on temperature profile after 5 ps IMPACT simulation. The maximum local heat flow is $2.2\times10^{15}$ W/cm\textsuperscript{2}}
\end{figure} 
Spline interpolation was used to increase the spatial resolution near the steep interface for the IMPACT simulations, helping both numerical stability and runtime due to needing a reduced number of nonlinear iterations. \color{black} For simplicity, the Coulomb logarithm was treated as a constant $\log \Lambda_\mathrm{ei} = \log \Lambda_\mathrm{ee}= 2.1484$. Note that in reality the plasma is only strongly coupled in the colder region of the gadolinium bubble beyond $\mathord{\sim}$\SI{1.7}{\milli\meter} and $\log\Lambda_\mathrm{ei}\approx8$ up to $\mathord{\sim}$\SI{1.6}{\milli\meter} in the hotter corona. \textcolor{black}{Reflective boundary conditions were used here as in the previous section and IMPACT used a timestep of \SI{1.334}{\femto\second}.} The $n_\mathrm{e}$ and $Z$ profiles did not evolve in the IMPACT simulation due to the treatment of the electric field discussed in section \ref{sec:vfp} that ensures quasineutrality  and the neglection of ion motion and ionisation models.

As with the KIPP simulations in the previous section,  there is an initial transient phase where the IMPACT heat flux gradually reduces from the Braginskii prediction as the distribution function rapidly moves away from Maxwellian.  Once again this transient phase is considered to be over when the ratio of the peak heat flow to the Braginskii prediction stops reducing. This ratio is not observed to change by more than 5\% after the first 5~ps of our 15.7~ps simulation. Therefore, we conclude that it safe to assume the transient phase is over after 5~ps, at which point the temperature front has advanced by approximately  8~$\mathrm{\upmu}$m leading to a \textcolor{black}{maximum temperature change of 41$\%$} as shown in Fig. \ref{fig:initVar}.

In comparing the IMPACT and SNB heat flow profiles we encountered an important subtlety concerning the implementation of the model\color{black}.\color{black}  While more recent publications concerning the SNB model\cite{SNBmag,Dario} use a formulation similar to that used here in section II C with separate electron-ion and electron-electron mfp's or collision frequencies, the original paper\cite{SNB} used a geometrically averaged mfp $\lambda_\mathrm{e} = \sqrt{Z\lambda_\mathrm{ei}^*\lambda_\mathrm{ei}}$. However, this averaging process is only valid for the case of homogeneous ionisation, and Fig. \ref{fig:Zvar} shows the large effect this has on the heat flow when the ionisation varies. While using separated mfp's provides a very good prediction of the preheat into the hohlraum, the peak heat flow to within 16\% and an improved estimate of the thermal conduction in the gas-fill region, the latter is still too large by a factor of $\mathord{\sim}2$. This discrepancy could potentially lead to an overestimate of hohlraum temperatures and thus cause issues similar to those arising with using an overly restrictive flux limiter\cite{HYDRA}.
\color{black}
\section{Discussion and Further Work}
\color{black}
The capability of the NFLF to closely match the results of EIC for the case of homogeneous density and ionisation is fairly impressive, considering that only 6 Lorentzians were needed for convergence compared to EIC's 64 moments (16,4 in the HL basis, chosen instead of LL as convergence is faster for this problem). This implies that the NFLF is about 5 times faster (assuming the NFLF's second-order ODE's take approximately twice the time to solve as EIC's first-order). However, this result should not be too surprising as both models are based on some kind of linearisation procedure, causing them to fail in almost exactly the same way for a nonlinear problem. For example, the lack of preheat or spatial shift in peak location predicted by the models are both features observed in the linear problem studied in section \ref{sec:linear}. The SNB model requires 25 groups for convergence resulting in an only slightly faster computation time \textcolor{black}{than} the EIC model.

Improving performance of the models for large temperature variations would require approaches that did not affect the desirable agreement in the linearised limit. For EIC, a simple method is nonlinear iteration; i.e., updating the right-hand side of equation (\ref{JiHeldFinal}) by adding on nonlinear terms such as $\displaystyle \frac{eE_\parallel}{m_\mathrm{e}}\frac{\partial \delta f}{\partial v_\parallel}-\sum_n A_n \frac{\partial \psi_n}{\partial s_\parallel}$ from the initial calculation and repeating until convergence. However, the computational time to apply the differential operators and separate into eigenvector components would probably increase the computational time by an undesirably large factor on the order of the number of moments used.

Conversely, a correct approach for improving the NFLF model is not immediately apparent, and probably requires deeper analysis of the link between the model and the VFP equation. However, it is conceivable that this could be done without additional computational expense; for example, replacing the $a^2 \lambda_\mathrm{ei}^2 \nabla^2$ term in equation (\ref{eq:NFLF}) with $a^2 (\lambda_\mathrm{ei}\nabla)\cdot (\lambda_\mathrm{ei}\nabla)$ would affect only nonlinear behaviour.

It is important to investigate sensitivity of divertor temperature to the errors in these models to confirm whether an accurate treatment of nonlocal transport can reconcile simulation and experiment.  Furthermore, the discrepancies observed with the SNB model when ionisation gradients are steep could potentially have critical knock-on effects for integrated \textcolor{black}{ICF} modeling; it needs to be determined whether further improvements to the SNB model are necessary to avoid this.

One key neglection in this work is the effect of electron-electron collisions on the anisotropic part of the distribution function $\bm{f}_1$ for the case of spatially varying ionisation.
It was shown in section \ref{sec:hydro} that inclusion of this in KIPP and EIC predicts a noticeably different nonlocal deviation (consider, for example, the value $b$) than would be predicted by using the phenomenological collision fix $\xi$ (which incorrectly predicts $b(Z)/b(\infty) = \kappa^\mathrm{(B)}(Z)/\kappa^\mathrm{(B)}(\infty)=\xi$ as depicted in  figure \ref{fig:bZ}).
But this did not seem to be the case for the more physically realistic large temperature variation studied in section \ref{sec:homo}, as using the value $r=2$ in the SNB model, derived in the linearised and Lorentz limits, seemed to be preferable to $r=3$.
Nevertheless, the use of $\xi$ in IMPACT as an ad-hoc substitution for a more complete approach to anisotropic electron-electron collisions could still potentially lead to inaccuracies in the heat flow predictions depicted in section \ref{sec:varion}, and this should be further investigated.

Less critical to our findings are the inaccuracies experienced by VFP codes in strongly coupled plasmas.
While this could play a role in the cooler part of the hohlraum wall studied in section \ref{sec:varion} where the Coulomb logarithm drops to $\mathord{\sim}2$ (theoretically rendering the effect of collisions in this region only accurate to $\mathord{\sim}50\%$) it does not affect the conclusion that the separated SNB model predicts the same heat flow into the wall as IMPACT while overpredicting that in the corona as both use the same treatment of $\log\Lambda$.
We have simply shown quantitatively that reduced models can be an effective stepping stone between hydrodynamic and VFP approaches.
However. this does act as a reminder that even a highly sophisticated VFP code could be faced with challenging inaccuracies in certain regions of the plasma (though it would surely still be an improvement to a purely hydrodynamic approach which would experience the same difficulties with strongly coupled plasmas); a potential method in overcoming this and incorporating large-angle collisions in a continuum code could be a Monte Carlo based approach\cite{Turrell}.
Similar points can be made for other deficiencies, such as collisions with neutrals and Fermi degeneracy, although these are probably slightly easier to address and incorporate into models\cite{VFPNeutrals,FermiDirac}.

Following on from these basic test problems and sensitivity tests, there are still important questions on predictive modeling of fusion plasma heat flows that could be answered using VFP codes. Firstly, the distribution function predicted by the SNB model should be compared to that of a fully kinetic code to assess the former's viability in predicting other transport coefficients or parameteric instabilities\cite{SherlockSNB}. Further modifications of the distribution function to a Dum-Langdon-Matte type super-Gaussian\cite{Dum1,*Dum2,Langdon,Matte} due to inverse bremsstrahlung by laser heating in inertial fusion could also significantly alter the transport processes\cite{InvBrem}. Furthermore, kinetic effects can still affect perpendicular transport (both heat flow and magnetic field advection rates) for moderate magnetisations\cite{reemerge,Archis}, this could be relevant to recent interest in magnetised hohlraums\cite{Montgomery,Strozzi2015} or magnetic islands in tokamaks\cite{Islands}; and while a few reduced models have been suggested to capture some of these aspects\cite{Dariomag,SNBmag} they still need to be properly validated with kinetic codes.
\color{black}
\section{Conclusions}
\color{black}
In conclusion, we have \textcolor{black}{compared three nonlocal models from ICF and MCF. We have demonstrated their} optimal implementations, revealing potential subtleties in the description of the models. We have demonstrated that the SNB model---using the original BGK operator, but scaled according to an analysis of small-amplitude temperature sinusoids ($r=2$), along with the modified source term $\nabla \cdot \bm{g}_1^{\mathrm{(mb)}}$ appearing on the right-hand side of equation (\ref{eq:SNB})---performs better than NFLF and EIC for the problems investigated with large temperature variations. Ensuring that the electron-electron and electron-ion collisionalities appear separately in this equation further improves agreement with VFP for a problem with spatially-varying ionisation. However, the problems studied with large temperature variation only reach a nonlocality parameter of $\mathord{\sim}15\%$, suggesting that SNB is most likely suitable for modeling hohlraum energetics problems (with the current exception of gas-fill heat flow, which is overestimated by a factor of $\mathord{\sim}2$) and mean SOL profiles but could break down at the even shorter scalelengths relevant to transient events. 

The NFLF and EIC models have been found to perform favourably against KIPP when predicting the rate of decay of a small-amplitude temperature pertubation over a wider range of collisionalities than the SNB.  However, these models overestimate the peak heat flux by up to $35\%$ in the case of a large temperature variation as well as failing to predict preheat. Additionally, a new analytic fit to kinetic results for temperature sinusoids has been presented in equation (\ref{eq:c1}) that could be useful in traditional Landau-fluid implementations. 

\section{Acknowledgements}
The authors would like to thank J Y Ji for sharing the numerical results of his closure\cite{JiClosure}, and A V Brantov for assistance in understanding his work with Bychenkov \emph{et al.}\cite{BrantovNoB,Brantov2013} \color{black} We would also like to express our gratitude towards A M Dimits, I Joseph, W Rozmus and L LoDestro for interesting discussions and pointing us towards a number of relevant references. We additionally appreciate valuable feedback on our manuscript from O Izacard and our Physics of Plasmas reviewer. Finally, thanks to M Zhao for sharing a script to analyse the KIPP results.

All data used to produce the figures in this work, along with other selected supporting data, can be found at \href{http://dx.doi.org/10.15124/3de4e00c-9087-4d95-89aa-9acd2071d3fb}{dx.doi.org/10.15124/3de4e00c-9087-4d95-89aa-9acd2071d3fb}.

This work is funded by EPSRC grants EP/K504178/1 and EP/M011372/1.
This work has been carried out within the framework of the EUROfusion Consortium and has received funding from the Euratom research and training programme 2014-2018 under grant agreement No 633053 (project reference CfP-AWP17-IFE-CCFE-01). The views and opinions expressed herein do not necessarily reflect those of the European Commission. This work was performed under the auspices of the U.S. Department of Energy by Lawrence Livermore National Laboratory under Contract DE-AC52-07NA27344. This work was supported by Royal Society award IE140365.
\appendix
\section{SNB in the hydrodynamic limit}
For long wavelength perturbations the diffusion term in equation \ref{eq:SNB} can be ignored and thus the distribution function and nonlocal heat flow easily computed in this limit. An outline of the derivation is given here, note that using the BGK collision operator and $\bm{g}_1^{(\textrm{mb})}$ in the source term gives the same $\delta f_0$ as when using the AWBS operator with $\bm{f}_1^{(\textrm{mb})}$ in the source term if $r=2$. 
The different consequences of choosing each source term $\bm{g}_1^{(\textrm{mb})}$, $\bm{f}_1^{(\textrm{mb})}$ are distinguished by the terms, on the left and right respectively, inside the curly brackets $\left\{\cdot,\cdot\right\}$.
\textcolor{black}{Note} that integration by parts is employed for the AWBS calculation and a change of variables to $u=v/\sqrt{2}v_\mathrm{T}$ is used. Additionally we define and use a further two dimensionless variables $X=\xi Z k^2\lambda_\mathrm{ei}(v)^2$ which is velocity-dependent and $X^{(\mathrm{B})}=\xi Z  k^2\lambda_\mathrm{ei}^{\mathrm{(B)}2}$ which is independent of velocity. \textcolor{black}{Numerical results of these calculations are summarised in Table \ref{tab:SNB}.}

\begin{widetext}
 {\begin{align*} 
 &\qquad \textbf{BGK} & & \qquad \textbf{AWBS} \\
   \delta f_0&=-\frac{\mathrm{i}Zk\lambda_\mathrm{ei}^{\mathrm(B)}}{r}\frac{\left\{\bm{g}_1^{(\textrm{mb})},\bm{f}_1^{(\textrm{mb})}\right\}}{3} &\delta f_0 &= \int_{\mathrlap{\infty}}^{\mathrlap{v}} \mathrm{d}v  \frac{ \mathrm{i}Zk\lambda_\mathrm{ei}(v)}{v}\frac{\left\{\bm{g}_1^{(\textrm{mb})},\bm{f}_1^{(\textrm{mb})}\right\}}{3} \\
   \delta \bm{Q} &= -\frac{2\uppi m_\mathrm{e}}{3}\int_{\mathrlap{0}}^{\mathrlap{\infty}}\mathrm{d}v \frac{X}{r}   \frac{v^5\left\{\bm{g}_1^{(\textrm{mb})},\bm{f}_1^{(\textrm{mb})}\right\}}{3}   &  \delta \bm{Q} &= -\frac{2\uppi m_\mathrm{e}}{3}\int_{\mathrlap{0}}^{\mathrlap{\infty}} \mathrm{d}v \frac{X}{10}   \frac{v^5\left\{\bm{g}_1^{(\textrm{mb})},\bm{f}_1^{(\textrm{mb})}\right\}}{3}   \\
    &= -\frac{32}{9\uppi}\int_{\mathrlap{0}}^{\mathrlap{\infty}} \mathrm{d} u \frac{u^{17}\{1,u^2-4\} \mathrm{e}^{-u^2}}{36}  \frac{X^{\textrm{(B)}}}{r}\bm{Q}^\mathrm{(B)}.  & &= -\frac{32}{9\uppi}\int_{\mathrlap{0}}^{\mathrlap{\infty}} \mathrm{d} u \frac{u^{17}\{1,u^2-4\} \mathrm{e}^{-u^2}}{360}  X^{\textrm{(B)}}\bm{Q}^\mathrm{(B)}.
     \end{align*}}
 \end{widetext}
 \section{Linearised SNB for arbitrary collisionality.}
 
A similar analyis can be performed with slightly greater difficulty at arbitrary collisionality. Integration by parts must be used again for the AWBS derivation, along with some mathematical identities. Recall that the electric field correction made by the SNB model is a nonlinear correction and does not come into play if the amplitude of the perturbation is infinitesimal:
\begin{widetext}
 {\begin{align*}
 &\qquad \textbf{BGK} & & \qquad \textbf{AWBS} \\
 \delta f_0&=-\frac{\mathrm{i}Zk\lambda_\mathrm{ei}^{\mathrm(B)}}{3r}\frac{\left\{\bm{g}_1^{(\textrm{mb})},\bm{f}_1^{(\textrm{mb})}\right\}}{1+X/3r} &\delta f_0 &= \mathrm{e}^{X/24}\int_{\mathrlap{\infty}}^{\mathrlap{v}} \mathrm{d}v \mathrm{e}^{-X/24}\frac{ \mathrm{i}Zk\lambda_\mathrm{ei}(v)}{v}\frac{\left\{\bm{g}_1^{(\textrm{mb})},\bm{f}_1^{(\textrm{mb})}\right\}}{3} \\
\delta \bm{Q} &= \frac{2\uppi m_\mathrm{e}}{3}\int_{\mathrlap{0}}^{\mathrlap{\infty}}\mathrm{d}v  \frac{v^5\left\{\bm{g}_1^{(\textrm{mb})},\bm{f}_1^{(\textrm{mb})}\right\}}{1+X/3r}  & \delta \bm{Q} &= -\frac{2\uppi m_\mathrm{e}}{3}\int_{\mathrlap{0}}^{\mathrlap{\infty}} \mathrm{d}v \frac{\gamma\left(\frac{5}{4},\frac{-X}{24}\right)\mathrm{e}^{-X/24}v^5\left\{\bm{g}_1^{(\textrm{mb})},\bm{f}_1^{(\textrm{mb})}\right\}}{(-X/24)^{1/4}}  \\
   \frac{ \bm{Q}}{\bm{Q}^\mathrm{(B)}}&= \int_{\mathrlap{0}}^{\mathrlap{\infty}} \mathrm{d}u \frac{u^{9}\{1,u^2-4\} \mathrm{e}^{-u^2}/12}{1+32X^{\textrm{(B)}}u^8/27\uppi r} .  &\frac{ \bm{Q}}{\bm{Q}^\mathrm{(B)}}
    &= \int_{\mathrlap{0}}^{\mathrlap{\infty}}\mathrm{d}u \frac{\gamma\left(\frac{1}{4},\frac{-4X^{\textrm{(B)}}u^8}{27\uppi}\right)\mathrm{e}^{-4X^{\textrm{(B)}}u^8/27\uppi}u^9\{1,u^2-4\} \mathrm{e}^{-u^2}}{12(-4 X^{\textrm{(B)}}u^8/27\uppi)^{1/4}},
     \end{align*}} 
\end{widetext}
where $\gamma$ is the incomplete gamma function. Computing the definite integral numerically with Mathematica shows that the AWBS heat flow can become negative for $X^\mathrm{(B)} > 0.0154$, which corresponds to $k\lambda_\mathrm{ei}^{\textrm{(B)}} > 0.124/\sqrt{\xi Z}$.


\begin{thebibliography}{80}%
\makeatletter
\providecommand \@ifxundefined [1]{%
 \@ifx{#1\undefined}
}%
\providecommand \@ifnum [1]{%
 \ifnum #1\expandafter \@firstoftwo
 \else \expandafter \@secondoftwo
 \fi
}%
\providecommand \@ifx [1]{%
 \ifx #1\expandafter \@firstoftwo
 \else \expandafter \@secondoftwo
 \fi
}%
\providecommand \natexlab [1]{#1}%
\providecommand \enquote  [1]{``#1''}%
\providecommand \bibnamefont  [1]{#1}%
\providecommand \bibfnamefont [1]{#1}%
\providecommand \citenamefont [1]{#1}%
\providecommand \href@noop [0]{\@secondoftwo}%
\providecommand \href [0]{\begingroup \@sanitize@url \@href}%
\providecommand \@href[1]{\@@startlink{#1}\@@href}%
\providecommand \@@href[1]{\endgroup#1\@@endlink}%
\providecommand \@sanitize@url [0]{\catcode `\\12\catcode `\$12\catcode
  `\&12\catcode `\#12\catcode `\^12\catcode `\_12\catcode `\%12\relax}%
\providecommand \@@startlink[1]{}%
\providecommand \@@endlink[0]{}%
\providecommand \url  [0]{\begingroup\@sanitize@url \@url }%
\providecommand \@url [1]{\endgroup\@href {#1}{\urlprefix }}%
\providecommand \urlprefix  [0]{URL }%
\providecommand \Eprint [0]{\href }%
\providecommand \doibase [0]{http://dx.doi.org/}%
\providecommand \selectlanguage [0]{\@gobble}%
\providecommand \bibinfo  [0]{\@secondoftwo}%
\providecommand \bibfield  [0]{\@secondoftwo}%
\providecommand \translation [1]{[#1]}%
\providecommand \BibitemOpen [0]{}%
\providecommand \bibitemStop [0]{}%
\providecommand \bibitemNoStop [0]{.\EOS\space}%
\providecommand \EOS [0]{\spacefactor3000\relax}%
\providecommand \BibitemShut  [1]{\csname bibitem#1\endcsname}%
\let\auto@bib@innerbib\@empty
\bibitem [{\citenamefont {Rosen}\ \emph {et~al.}(2011)\citenamefont {Rosen},
  \citenamefont {Scott}, \citenamefont {Hinkel}, \citenamefont {Williams},
  \citenamefont {Callahan}, \citenamefont {Town}, \citenamefont {Divol},
  \citenamefont {Michel}, \citenamefont {Kruer}, \citenamefont {Suter} \emph
  {et~al.}}]{HYDRA}%
  \BibitemOpen
  \bibfield  {author} {\bibinfo {author} {\bibfnamefont {M.~D.}\ \bibnamefont
  {Rosen}}, \bibinfo {author} {\bibfnamefont {H.~A.}\ \bibnamefont {Scott}},
  \bibinfo {author} {\bibfnamefont {D.~E.}\ \bibnamefont {Hinkel}}, \bibinfo
  {author} {\bibfnamefont {E.~A.}\ \bibnamefont {Williams}}, \bibinfo {author}
  {\bibfnamefont {D.~A.}\ \bibnamefont {Callahan}}, \bibinfo {author}
  {\bibfnamefont {R.~P.~J.}\ \bibnamefont {Town}}, \bibinfo {author}
  {\bibfnamefont {L.}~\bibnamefont {Divol}}, \bibinfo {author} {\bibfnamefont
  {P.~A.}\ \bibnamefont {Michel}}, \bibinfo {author} {\bibfnamefont {W.~L.}\
  \bibnamefont {Kruer}}, \bibinfo {author} {\bibfnamefont {L.~J.}\ \bibnamefont
  {Suter}},  \emph {et~al.},\ }\href {\doibase
  https://doi.org/10.1016/j.hedp.2011.03.008} {\bibfield  {journal} {\bibinfo
  {journal} {HEDP}\ }\textbf {\bibinfo {volume} {7}},\ \bibinfo {pages} {180}
  (\bibinfo {year} {2011})}\BibitemShut {NoStop}%
\bibitem [{\citenamefont {Dudson}\ \emph {et~al.}(2009)\citenamefont {Dudson},
  \citenamefont {Umansky}, \citenamefont {Xu}, \citenamefont {Snyder},\ and\
  \citenamefont {Wilson}}]{BOUTframework}%
  \BibitemOpen
  \bibfield  {author} {\bibinfo {author} {\bibfnamefont {B.~D.}\ \bibnamefont
  {Dudson}}, \bibinfo {author} {\bibfnamefont {M.~V.}\ \bibnamefont {Umansky}},
  \bibinfo {author} {\bibfnamefont {X.~Q.}\ \bibnamefont {Xu}}, \bibinfo
  {author} {\bibfnamefont {P.~B.}\ \bibnamefont {Snyder}}, \ and\ \bibinfo
  {author} {\bibfnamefont {H.~R.}\ \bibnamefont {Wilson}},\ }\href {\doibase
  http://dx.doi.org/10.1016/j.cpc.2009.03.008} {\bibfield  {journal} {\bibinfo
  {journal} {Computer Physics Communications}\ }\textbf {\bibinfo {volume}
  {180}},\ \bibinfo {pages} {1467 } (\bibinfo {year} {2009})}\BibitemShut
  {NoStop}%
\bibitem [{\citenamefont {Ji}, \citenamefont {Held},\ and\ \citenamefont
  {Sovinec}(2009)}]{JiHeld}%
  \BibitemOpen
  \bibfield  {author} {\bibinfo {author} {\bibfnamefont {J.~Y.}\ \bibnamefont
  {Ji}}, \bibinfo {author} {\bibfnamefont {E.~D.}\ \bibnamefont {Held}}, \ and\
  \bibinfo {author} {\bibfnamefont {C.~R.}\ \bibnamefont {Sovinec}},\
  }\href@noop {} {\bibfield  {journal} {\bibinfo  {journal} {Phys. Plasmas}\
  }\textbf {\bibinfo {volume} {16}},\ \bibinfo {pages} {022312} (\bibinfo
  {year} {2009})}\BibitemShut {NoStop}%
\bibitem [{\citenamefont {Omotani}\ and\ \citenamefont
  {Dudson}(2013)}]{Omotani}%
  \BibitemOpen
  \bibfield  {author} {\bibinfo {author} {\bibfnamefont {J.~T.}\ \bibnamefont
  {Omotani}}\ and\ \bibinfo {author} {\bibfnamefont {B.~D.}\ \bibnamefont
  {Dudson}},\ }\href {http://stacks.iop.org/0741-3335/55/i=5/a=055009}
  {\bibfield  {journal} {\bibinfo  {journal} {Plasma Phys. Control. Fusion}\
  }\textbf {\bibinfo {volume} {55}},\ \bibinfo {pages} {055009} (\bibinfo
  {year} {2013})}\BibitemShut {NoStop}%
\bibitem [{\citenamefont {Omotani}\ \emph {et~al.}(2015)\citenamefont
  {Omotani}, \citenamefont {Dudson}, \citenamefont {Havl\'{i}\v{c}kov\'{a}},\
  and\ \citenamefont {Umansky}}]{Hermite}%
  \BibitemOpen
  \bibfield  {author} {\bibinfo {author} {\bibfnamefont {J.~T.}\ \bibnamefont
  {Omotani}}, \bibinfo {author} {\bibfnamefont {B.~D.}\ \bibnamefont {Dudson}},
  \bibinfo {author} {\bibfnamefont {E.}~\bibnamefont {Havl\'{i}\v{c}kov\'{a}}},
  \ and\ \bibinfo {author} {\bibfnamefont {M.}~\bibnamefont {Umansky}},\ }\href
  {\doibase http://dx.doi.org/10.1016/j.jnucmat.2014.10.040} {\bibfield
  {journal} {\bibinfo  {journal} {J. Nucl. Mater.}\ }\textbf {\bibinfo {volume}
  {463}},\ \bibinfo {pages} {769 } (\bibinfo {year} {2015})}\BibitemShut
  {NoStop}%
\bibitem [{\citenamefont {Dimits}, \citenamefont {Joseph},\ and\ \citenamefont
  {Umansky}(2013)}]{NFLFTheory}%
  \BibitemOpen
  \bibfield  {author} {\bibinfo {author} {\bibfnamefont {A.~M.}\ \bibnamefont
  {Dimits}}, \bibinfo {author} {\bibfnamefont {I.}~\bibnamefont {Joseph}}, \
  and\ \bibinfo {author} {\bibfnamefont {M.~V.}\ \bibnamefont {Umansky}},\
  }\href@noop {} {\bibfield  {journal} {\bibinfo  {journal} {Bull. American
  Phys. Soc.}\ }\textbf {\bibinfo {volume} {58}},\ \bibinfo {pages} {281}
  (\bibinfo {year} {2013})}\BibitemShut {NoStop}%
\bibitem [{\citenamefont {Dimits}, \citenamefont {Joseph},\ and\ \citenamefont
  {Umansky}(2014)}]{NFLF2014}%
  \BibitemOpen
  \bibfield  {author} {\bibinfo {author} {\bibfnamefont {A.~M.}\ \bibnamefont
  {Dimits}}, \bibinfo {author} {\bibfnamefont {I.}~\bibnamefont {Joseph}}, \
  and\ \bibinfo {author} {\bibfnamefont {M.~V.}\ \bibnamefont {Umansky}},\
  }\href {\doibase 10.1063/1.4876617} {\bibfield  {journal} {\bibinfo
  {journal} {Phys. Plasmas}\ }\textbf {\bibinfo {volume} {21}},\ \bibinfo
  {pages} {055907} (\bibinfo {year} {2014})}\BibitemShut {NoStop}%
\bibitem [{\citenamefont {Umansky}\ \emph {et~al.}(2015)\citenamefont
  {Umansky}, \citenamefont {Dimits}, \citenamefont {Joseph}, \citenamefont
  {Omotani},\ and\ \citenamefont {Rognlien}}]{NFLFImplementation}%
  \BibitemOpen
  \bibfield  {author} {\bibinfo {author} {\bibfnamefont {M.~V.}\ \bibnamefont
  {Umansky}}, \bibinfo {author} {\bibfnamefont {A.~M.}\ \bibnamefont {Dimits}},
  \bibinfo {author} {\bibfnamefont {I.}~\bibnamefont {Joseph}}, \bibinfo
  {author} {\bibfnamefont {J.~T.}\ \bibnamefont {Omotani}}, \ and\ \bibinfo
  {author} {\bibfnamefont {T.~D.}\ \bibnamefont {Rognlien}},\ }\href {\doibase
  http://dx.doi.org/10.1016/j.jnucmat.2014.10.015} {\bibfield  {journal}
  {\bibinfo  {journal} {J. Nucl. Mater.}\ }\textbf {\bibinfo {volume} {463}},\
  \bibinfo {pages} {506} (\bibinfo {year} {2015})},\ \bibinfo {note}
  {{\relax{}Proceedings} of the 21st International Conference on Plasma-Surface
  Interactions in Controlled Fusion Devices}\BibitemShut {NoStop}%
\bibitem [{\citenamefont {Schurtz}, \citenamefont {Nicola{\"i}},\ and\
  \citenamefont {Busquet}(2000)}]{SNB}%
  \BibitemOpen
  \bibfield  {author} {\bibinfo {author} {\bibfnamefont {G.~P.}\ \bibnamefont
  {Schurtz}}, \bibinfo {author} {\bibfnamefont {{\relax Ph}.~D.}\ \bibnamefont
  {Nicola{\"i}}}, \ and\ \bibinfo {author} {\bibfnamefont {M.}~\bibnamefont
  {Busquet}},\ }\href {\doibase 10.1063/1.1289512} {\bibfield  {journal}
  {\bibinfo  {journal} {Phys. Plasmas}\ }\textbf {\bibinfo {volume} {7}},\
  \bibinfo {pages} {4238} (\bibinfo {year} {2000})}\BibitemShut {NoStop}%
\bibitem [{\citenamefont {Nicola{\"i}}, \citenamefont {Feugeas},\ and\
  \citenamefont {Schurtz}(2006)}]{SNBmag}%
  \BibitemOpen
  \bibfield  {author} {\bibinfo {author} {\bibfnamefont {{\relax Ph}.~D.}\
  \bibnamefont {Nicola{\"i}}}, \bibinfo {author} {\bibfnamefont {{\relax
  J.-L}.~A.}\ \bibnamefont {Feugeas}}, \ and\ \bibinfo {author} {\bibfnamefont
  {G.~P.}\ \bibnamefont {Schurtz}},\ }\href {\doibase 10.1063/1.2179392}
  {\bibfield  {journal} {\bibinfo  {journal} {Phys. Plasmas}\ }\textbf
  {\bibinfo {volume} {13}},\ \bibinfo {pages} {032701} (\bibinfo {year}
  {2006})}\BibitemShut {NoStop}%
\bibitem [{\citenamefont {Del~Sorbo}\ \emph {et~al.}(2015)\citenamefont
  {Del~Sorbo}, \citenamefont {Feugeas}, \citenamefont {Nicola\"{i}},
  \citenamefont {Olazabal-Loum\'{e}}, \citenamefont {Dubroca}, \citenamefont
  {Guisset}, \citenamefont {Touati},\ and\ \citenamefont {Tikhonchuk}}]{Dario}%
  \BibitemOpen
  \bibfield  {author} {\bibinfo {author} {\bibfnamefont {D.}~\bibnamefont
  {Del~Sorbo}}, \bibinfo {author} {\bibfnamefont {{\relax J.-L.}.}~\bibnamefont
  {Feugeas}}, \bibinfo {author} {\bibfnamefont {{\relax Ph}.}~\bibnamefont
  {Nicola\"{i}}}, \bibinfo {author} {\bibfnamefont {M.}~\bibnamefont
  {Olazabal-Loum\'{e}}}, \bibinfo {author} {\bibfnamefont {B.}~\bibnamefont
  {Dubroca}}, \bibinfo {author} {\bibfnamefont {S.}~\bibnamefont {Guisset}},
  \bibinfo {author} {\bibfnamefont {M.}~\bibnamefont {Touati}}, \ and\ \bibinfo
  {author} {\bibfnamefont {V.}~\bibnamefont {Tikhonchuk}},\ }\href {\doibase
  10.1063/1.4926824} {\bibfield  {journal} {\bibinfo  {journal} {Phys.
  Plasmas}\ }\textbf {\bibinfo {volume} {22}},\ \bibinfo {pages} {082706}
  (\bibinfo {year} {2015})}\BibitemShut {NoStop}%
\bibitem [{\citenamefont {Del~Sorbo}\ \emph {et~al.}(2016)\citenamefont
  {Del~Sorbo}, \citenamefont {Feugeas}, \citenamefont {Nicola\"{i}},
  \citenamefont {Olazabal-Loum\'{e}}, \citenamefont {Dubroca},\ and\
  \citenamefont {Tikhonchuk}}]{Dariomag}%
  \BibitemOpen
  \bibfield  {author} {\bibinfo {author} {\bibfnamefont {D.}~\bibnamefont
  {Del~Sorbo}}, \bibinfo {author} {\bibfnamefont {J.-L.}\ \bibnamefont
  {Feugeas}}, \bibinfo {author} {\bibfnamefont {P.}~\bibnamefont
  {Nicola\"{i}}}, \bibinfo {author} {\bibfnamefont {M.}~\bibnamefont
  {Olazabal-Loum\'{e}}}, \bibinfo {author} {\bibfnamefont {B.}~\bibnamefont
  {Dubroca}}, \ and\ \bibinfo {author} {\bibfnamefont {V.}~\bibnamefont
  {Tikhonchuk}},\ }\href {\doibase 10.1017/S0263034616000252} {\bibfield
  {journal} {\bibinfo  {journal} {Laser and Particle Beams}\ }\textbf {\bibinfo
  {volume} {34}},\ \bibinfo {pages} {412–} (\bibinfo {year}
  {2016})}\BibitemShut {NoStop}%
\bibitem [{\citenamefont {Cao}, \citenamefont {Moses},\ and\ \citenamefont
  {Delettrez}(2015)}]{Cao}%
  \BibitemOpen
  \bibfield  {author} {\bibinfo {author} {\bibfnamefont {D.}~\bibnamefont
  {Cao}}, \bibinfo {author} {\bibfnamefont {G.}~\bibnamefont {Moses}}, \ and\
  \bibinfo {author} {\bibfnamefont {J.}~\bibnamefont {Delettrez}},\ }\href
  {\doibase 10.1063/1.4928445} {\bibfield  {journal} {\bibinfo  {journal}
  {Phys. Plasmas}\ }\textbf {\bibinfo {volume} {22}},\ \bibinfo {pages}
  {082308} (\bibinfo {year} {2015})}\BibitemShut {NoStop}%
\bibitem [{\citenamefont {Braginskii}(1965)}]{Braginskii}%
  \BibitemOpen
  \bibfield  {author} {\bibinfo {author} {\bibfnamefont {S.~I.}\ \bibnamefont
  {Braginskii}},\ }\enquote {\bibinfo {title} {Transport processes in a
  plasma},}\ in\ \href@noop {} {\emph {\bibinfo {booktitle} {Reviews of Plasma
  Physics}}},\ Vol.~\bibinfo {volume} {1},\ \bibinfo {editor} {edited by\
  \bibinfo {editor} {\bibfnamefont {A.~M.}\ \bibnamefont {Leontovich}}}\
  (\bibinfo  {publisher} {Consultants Bureau},\ \bibinfo {address} {New York},\
  \bibinfo {year} {1965})\ p.\ \bibinfo {pages} {251}\BibitemShut {NoStop}%
\bibitem [{\citenamefont {Epperlein}\ and\ \citenamefont
  {Haines}(1986)}]{EppHaines}%
  \BibitemOpen
  \bibfield  {author} {\bibinfo {author} {\bibfnamefont {E.~M.}\ \bibnamefont
  {Epperlein}}\ and\ \bibinfo {author} {\bibfnamefont {M.~G.}\ \bibnamefont
  {Haines}},\ }\href {\doibase 10.1063/1.865901} {\bibfield  {journal}
  {\bibinfo  {journal} {Phys. Fluids}\ }\textbf {\bibinfo {volume} {29}},\
  \bibinfo {pages} {1029} (\bibinfo {year} {1986})}\BibitemShut {NoStop}%
\bibitem [{\citenamefont {Epperlein}\ and\ \citenamefont
  {Short}(1991)}]{EppShort}%
  \BibitemOpen
  \bibfield  {author} {\bibinfo {author} {\bibfnamefont {E.~M.}\ \bibnamefont
  {Epperlein}}\ and\ \bibinfo {author} {\bibfnamefont {R.~W.}\ \bibnamefont
  {Short}},\ }\href {\doibase 10.1063/1.859789} {\bibfield  {journal} {\bibinfo
   {journal} {Phys. Fluids B}\ }\textbf {\bibinfo {volume} {3}},\ \bibinfo
  {pages} {3092} (\bibinfo {year} {1991})}\BibitemShut {NoStop}%
\bibitem [{\citenamefont {Chodura}(1990)}]{Chodura}%
  \BibitemOpen
  \bibfield  {author} {\bibinfo {author} {\bibfnamefont {R.}~\bibnamefont
  {Chodura}},\ }\href {\doibase 10.1002/ctpp.2150300125} {\bibfield  {journal}
  {\bibinfo  {journal} {Contrib. Plasma Phys.}\ }\textbf {\bibinfo {volume}
  {30}},\ \bibinfo {pages} {153} (\bibinfo {year} {1990})}\BibitemShut
  {NoStop}%
\bibitem [{\citenamefont {Batishchev}\ \emph {et~al.}(1997)\citenamefont
  {Batishchev}, \citenamefont {Krasheninnikov}, \citenamefont {Catto},
  \citenamefont {Batishcheva}, \citenamefont {Sigmar}, \citenamefont {Xu},
  \citenamefont {Byers}, \citenamefont {Rognlien}, \citenamefont {Cohen},
  \citenamefont {Shoucri},\ and\ \citenamefont {Shkarofskii}}]{Batishchev}%
  \BibitemOpen
  \bibfield  {author} {\bibinfo {author} {\bibfnamefont {O.~V.}\ \bibnamefont
  {Batishchev}}, \bibinfo {author} {\bibfnamefont {S.~I.}\ \bibnamefont
  {Krasheninnikov}}, \bibinfo {author} {\bibfnamefont {P.~J.}\ \bibnamefont
  {Catto}}, \bibinfo {author} {\bibfnamefont {A.~A.}\ \bibnamefont
  {Batishcheva}}, \bibinfo {author} {\bibfnamefont {D.~J.}\ \bibnamefont
  {Sigmar}}, \bibinfo {author} {\bibfnamefont {X.~Q.}\ \bibnamefont {Xu}},
  \bibinfo {author} {\bibfnamefont {J.~A.}\ \bibnamefont {Byers}}, \bibinfo
  {author} {\bibfnamefont {T.~D.}\ \bibnamefont {Rognlien}}, \bibinfo {author}
  {\bibfnamefont {R.~H.}\ \bibnamefont {Cohen}}, \bibinfo {author}
  {\bibfnamefont {M.~M.}\ \bibnamefont {Shoucri}}, \ and\ \bibinfo {author}
  {\bibfnamefont {I.~P.}\ \bibnamefont {Shkarofskii}},\ }\href {\doibase
  10.1063/1.872280} {\bibfield  {journal} {\bibinfo  {journal} {Phys. Plasmas}\
  }\textbf {\bibinfo {volume} {4}},\ \bibinfo {pages} {1672} (\bibinfo {year}
  {1997})}\BibitemShut {NoStop}%
\bibitem [{\citenamefont {Schneider}\ \emph {et~al.}(1992)\citenamefont
  {Schneider}, \citenamefont {Reiter}, \citenamefont {Zehrfeld}, \citenamefont
  {Braams}, \citenamefont {Baelmans}, \citenamefont {Geiger}, \citenamefont
  {Kastelewicz}, \citenamefont {Neuhauser},\ and\ \citenamefont
  {Wunderlich}}]{SOLPS1}%
  \BibitemOpen
  \bibfield  {author} {\bibinfo {author} {\bibfnamefont {R.}~\bibnamefont
  {Schneider}}, \bibinfo {author} {\bibfnamefont {D.}~\bibnamefont {Reiter}},
  \bibinfo {author} {\bibfnamefont {H.~P.}\ \bibnamefont {Zehrfeld}}, \bibinfo
  {author} {\bibfnamefont {B.}~\bibnamefont {Braams}}, \bibinfo {author}
  {\bibfnamefont {M.}~\bibnamefont {Baelmans}}, \bibinfo {author}
  {\bibfnamefont {J.}~\bibnamefont {Geiger}}, \bibinfo {author} {\bibfnamefont
  {H.}~\bibnamefont {Kastelewicz}}, \bibinfo {author} {\bibfnamefont
  {J.}~\bibnamefont {Neuhauser}}, \ and\ \bibinfo {author} {\bibfnamefont
  {R.}~\bibnamefont {Wunderlich}},\ }\href {\doibase
  http://dx.doi.org/10.1016/S0022-3115(06)80147-9} {\bibfield  {journal}
  {\bibinfo  {journal} {J. Nucl. Mater.}\ }\textbf {\bibinfo {volume}
  {196--198}},\ \bibinfo {pages} {810} (\bibinfo {year} {1992})}\BibitemShut
  {NoStop}%
\bibitem [{\citenamefont {Reiter}(1992)}]{SOLPS2}%
  \BibitemOpen
  \bibfield  {author} {\bibinfo {author} {\bibfnamefont {D.}~\bibnamefont
  {Reiter}},\ }\href {\doibase http://dx.doi.org/10.1016/S0022-3115(06)80014-0}
  {\bibfield  {journal} {\bibinfo  {journal} {J. Nucl. Mater.}\ }\textbf
  {\bibinfo {volume} {196--198}},\ \bibinfo {pages} {80} (\bibinfo {year}
  {1992})}\BibitemShut {NoStop}%
\bibitem [{\citenamefont {Rognlien}\ \emph {et~al.}(1994)\citenamefont
  {Rognlien}, \citenamefont {Brown}, \citenamefont {Campbell}, \citenamefont
  {Kaiser}, \citenamefont {Knoll}, \citenamefont {\relax{McHugh}},
  \citenamefont {Porter}, \citenamefont {Rensink},\ and\ \citenamefont
  {Smith}}]{UEDGE}%
  \BibitemOpen
  \bibfield  {author} {\bibinfo {author} {\bibfnamefont {T.~D.}\ \bibnamefont
  {Rognlien}}, \bibinfo {author} {\bibfnamefont {P.~N.}\ \bibnamefont {Brown}},
  \bibinfo {author} {\bibfnamefont {R.~B.}\ \bibnamefont {Campbell}}, \bibinfo
  {author} {\bibfnamefont {T.~B.}\ \bibnamefont {Kaiser}}, \bibinfo {author}
  {\bibfnamefont {D.~A.}\ \bibnamefont {Knoll}}, \bibinfo {author}
  {\bibfnamefont {P.~R.}\ \bibnamefont {\relax{McHugh}}}, \bibinfo {author}
  {\bibfnamefont {G.~D.}\ \bibnamefont {Porter}}, \bibinfo {author}
  {\bibfnamefont {M.~E.}\ \bibnamefont {Rensink}}, \ and\ \bibinfo {author}
  {\bibfnamefont {G.~R.}\ \bibnamefont {Smith}},\ }\href {\doibase
  10.1002/ctpp.2150340241} {\bibfield  {journal} {\bibinfo  {journal} {Contrib.
  Plasma Phys.}\ }\textbf {\bibinfo {volume} {34}},\ \bibinfo {pages} {362}
  (\bibinfo {year} {1994})}\BibitemShut {NoStop}%
\bibitem [{\citenamefont {Chankin}\ and\ \citenamefont
  {Coster}(2009)}]{Chankin2D}%
  \BibitemOpen
  \bibfield  {author} {\bibinfo {author} {\bibfnamefont {A.~V.}\ \bibnamefont
  {Chankin}}\ and\ \bibinfo {author} {\bibfnamefont {D.~P.}\ \bibnamefont
  {Coster}},\ }\href {\doibase http://dx.doi.org/10.1016/j.jnucmat.2009.01.307}
  {\bibfield  {journal} {\bibinfo  {journal} {J. Nucl. Mater.}\ }\textbf
  {\bibinfo {volume} {390-391}},\ \bibinfo {pages} {319} (\bibinfo {year}
  {2009})},\ \bibinfo {note} {{Proceedings} of the 18th International
  Conference on Plasma-Surface Interactions in Controlled Fusion
  Devices}\BibitemShut {NoStop}%
\bibitem [{\citenamefont {Churchill}\ \emph {et~al.}(2016)\citenamefont
  {Churchill}, \citenamefont {Canik}, \citenamefont {Chang}, \citenamefont
  {Hager}, \citenamefont {Leonard}, \citenamefont {Maingi}, \citenamefont
  {Nazikian},\ and\ \citenamefont {Stotler}}]{Churchill}%
  \BibitemOpen
  \bibfield  {author} {\bibinfo {author} {\bibfnamefont {R.~M.}\ \bibnamefont
  {Churchill}}, \bibinfo {author} {\bibfnamefont {J.~M.}\ \bibnamefont
  {Canik}}, \bibinfo {author} {\bibfnamefont {C.~S.}\ \bibnamefont {Chang}},
  \bibinfo {author} {\bibfnamefont {R.}~\bibnamefont {Hager}}, \bibinfo
  {author} {\bibfnamefont {A.~W.}\ \bibnamefont {Leonard}}, \bibinfo {author}
  {\bibfnamefont {R.}~\bibnamefont {Maingi}}, \bibinfo {author} {\bibfnamefont
  {R.}~\bibnamefont {Nazikian}}, \ and\ \bibinfo {author} {\bibfnamefont
  {D.~P.}\ \bibnamefont {Stotler}},\ }\href {\doibase
  http://dx.doi.org/10.1016/j.nme.2016.12.013} {\bibfield  {journal} {\bibinfo
  {journal} {Nuclear Materials and Energy}\ } (\bibinfo {year} {2016}),\
  http://dx.doi.org/10.1016/j.nme.2016.12.013}\BibitemShut {NoStop}%
\bibitem [{\citenamefont {Horacek}\ \emph {et~al.}(2003)\citenamefont
  {Horacek}, \citenamefont {Pitts}, \citenamefont {Stangeby}, \citenamefont
  {Batishchev},\ and\ \citenamefont {Loarte}}]{Horacek2003}%
  \BibitemOpen
  \bibfield  {author} {\bibinfo {author} {\bibfnamefont {J.}~\bibnamefont
  {Horacek}}, \bibinfo {author} {\bibfnamefont {R.~A.}\ \bibnamefont {Pitts}},
  \bibinfo {author} {\bibfnamefont {P.~C.}\ \bibnamefont {Stangeby}}, \bibinfo
  {author} {\bibfnamefont {O.}~\bibnamefont {Batishchev}}, \ and\ \bibinfo
  {author} {\bibfnamefont {A.}~\bibnamefont {Loarte}},\ }\href {\doibase
  http://dx.doi.org/10.1016/S0022-3115(02)01479-4} {\bibfield  {journal}
  {\bibinfo  {journal} {J Nucl. Mater.}\ }\textbf {\bibinfo {volume} {313}},\
  \bibinfo {pages} {931} (\bibinfo {year} {2003})}\BibitemShut {NoStop}%
\bibitem [{\citenamefont {Jaworski}\ \emph {et~al.}(2012)\citenamefont
  {Jaworski}, \citenamefont {Bell}, \citenamefont {Gray}, \citenamefont
  {Kaita}, \citenamefont {Kallman}, \citenamefont {Kugel}, \citenamefont
  {LeBlanc}, \citenamefont {McLean}, \citenamefont {Sabbagh}, \citenamefont
  {Soukhanovskii}, \citenamefont {Stotler},\ and\ \citenamefont
  {Surla}}]{JAWORSKI2012}%
  \BibitemOpen
  \bibfield  {author} {\bibinfo {author} {\bibfnamefont {M.~A.}\ \bibnamefont
  {Jaworski}}, \bibinfo {author} {\bibfnamefont {M.~G.}\ \bibnamefont {Bell}},
  \bibinfo {author} {\bibfnamefont {T.~K.}\ \bibnamefont {Gray}}, \bibinfo
  {author} {\bibfnamefont {R.}~\bibnamefont {Kaita}}, \bibinfo {author}
  {\bibfnamefont {J.}~\bibnamefont {Kallman}}, \bibinfo {author} {\bibfnamefont
  {H.~W.}\ \bibnamefont {Kugel}}, \bibinfo {author} {\bibfnamefont
  {B.}~\bibnamefont {LeBlanc}}, \bibinfo {author} {\bibfnamefont {A.~G.}\
  \bibnamefont {McLean}}, \bibinfo {author} {\bibfnamefont {S.~A.}\
  \bibnamefont {Sabbagh}}, \bibinfo {author} {\bibfnamefont {V.~A.}\
  \bibnamefont {Soukhanovskii}}, \bibinfo {author} {\bibfnamefont {D.~P.}\
  \bibnamefont {Stotler}}, \ and\ \bibinfo {author} {\bibfnamefont
  {V.}~\bibnamefont {Surla}},\ }\href {\doibase
  http://dx.doi.org/10.1016/j.fusengdes.2011.07.013} {\bibfield  {journal}
  {\bibinfo  {journal} {Fusion Eng. and Design}\ }\textbf {\bibinfo {volume}
  {87}},\ \bibinfo {pages} {1711 } (\bibinfo {year} {2012})},\ \bibinfo {note}
  {the 2nd International Symposium of Lithium Application for Fusion
  Devices}\BibitemShut {NoStop}%
\bibitem [{\citenamefont {Jaworski}\ \emph {et~al.}(2013)\citenamefont
  {Jaworski}, \citenamefont {Bell}, \citenamefont {Gray}, \citenamefont
  {Kaita}, \citenamefont {Kaganovich}, \citenamefont {Kallman}, \citenamefont
  {Kugel}, \citenamefont {LeBlanc}, \citenamefont {McLean}, \citenamefont
  {Sabbagh}, \citenamefont {Scotti}, \citenamefont {Soukhanovskii},\ and\
  \citenamefont {Stotler}}]{JAWORSKI2013}%
  \BibitemOpen
  \bibfield  {author} {\bibinfo {author} {\bibfnamefont {M.~A.}\ \bibnamefont
  {Jaworski}}, \bibinfo {author} {\bibfnamefont {M.~G.}\ \bibnamefont {Bell}},
  \bibinfo {author} {\bibfnamefont {T.~K.}\ \bibnamefont {Gray}}, \bibinfo
  {author} {\bibfnamefont {R.}~\bibnamefont {Kaita}}, \bibinfo {author}
  {\bibfnamefont {I.}~\bibnamefont {Kaganovich}}, \bibinfo {author}
  {\bibfnamefont {J.}~\bibnamefont {Kallman}}, \bibinfo {author} {\bibfnamefont
  {H.~W.}\ \bibnamefont {Kugel}}, \bibinfo {author} {\bibfnamefont
  {B.}~\bibnamefont {LeBlanc}}, \bibinfo {author} {\bibfnamefont {A.~G.}\
  \bibnamefont {McLean}}, \bibinfo {author} {\bibfnamefont {S.~A.}\
  \bibnamefont {Sabbagh}}, \bibinfo {author} {\bibfnamefont {F.}~\bibnamefont
  {Scotti}}, \bibinfo {author} {\bibfnamefont {V.~A.}\ \bibnamefont
  {Soukhanovskii}}, \ and\ \bibinfo {author} {\bibfnamefont {D.~P.}\
  \bibnamefont {Stotler}},\ }\href {\doibase
  http://dx.doi.org/10.1016/j.jnucmat.2013.01.076} {\bibfield  {journal}
  {\bibinfo  {journal} {J. Nucl. Mater.}\ }\textbf {\bibinfo {volume} {438}},\
  \bibinfo {pages} {S384 } (\bibinfo {year} {2013})},\ \bibinfo {note} {{\relax
  Proceedings }of the 20th International Conference on Plasma-Surface
  Interactions in Controlled Fusion Devices}\BibitemShut {NoStop}%
\bibitem [{\citenamefont {Izacard}(2016)}]{Izacard}%
  \BibitemOpen
  \bibfield  {author} {\bibinfo {author} {\bibfnamefont {O.}~\bibnamefont
  {Izacard}},\ }\href {\doibase 10.1063/1.4960123} {\bibfield  {journal}
  {\bibinfo  {journal} {Phys. Plasmas}\ }\textbf {\bibinfo {volume} {23}},\
  \bibinfo {pages} {082504} (\bibinfo {year} {2016})}\BibitemShut {NoStop}%
\bibitem [{\citenamefont {{\v{D}}uran}\ \emph {et~al.}(2015)\citenamefont
  {{\v{D}}uran}, \citenamefont {Je{\v{s}}ko}, \citenamefont {Fuchs},
  \citenamefont {Groth}, \citenamefont {Guillemaut}, \citenamefont {Gunn},
  \citenamefont {Horacek}, \citenamefont {Pitts}, \citenamefont {Tskhakaya},\
  and\ \citenamefont {Contributors}}]{Duran2015}%
  \BibitemOpen
  \bibfield  {author} {\bibinfo {author} {\bibfnamefont {I.}~\bibnamefont
  {{\v{D}}uran}}, \bibinfo {author} {\bibfnamefont {K.}~\bibnamefont
  {Je{\v{s}}ko}}, \bibinfo {author} {\bibfnamefont {V.}~\bibnamefont {Fuchs}},
  \bibinfo {author} {\bibfnamefont {M.}~\bibnamefont {Groth}}, \bibinfo
  {author} {\bibfnamefont {C.}~\bibnamefont {Guillemaut}}, \bibinfo {author}
  {\bibfnamefont {J.~P.}\ \bibnamefont {Gunn}}, \bibinfo {author}
  {\bibfnamefont {J.}~\bibnamefont {Horacek}}, \bibinfo {author} {\bibfnamefont
  {R.~A.}\ \bibnamefont {Pitts}}, \bibinfo {author} {\bibfnamefont
  {D.}~\bibnamefont {Tskhakaya}}, \ and\ \bibinfo {author} {\bibfnamefont
  {J.-E.}\ \bibnamefont {Contributors}},\ }\href {\doibase
  http://dx.doi.org/10.1016/j.jnucmat.2015.01.051} {\bibfield  {journal}
  {\bibinfo  {journal} {J. Nucl. Mater.}\ }\textbf {\bibinfo {volume} {463}},\
  \bibinfo {pages} {432} (\bibinfo {year} {2015})}\BibitemShut {NoStop}%
\bibitem [{\citenamefont {Turnyanskiy}\ \emph {et~al.}(2015)\citenamefont
  {Turnyanskiy}, \citenamefont {Neu}, \citenamefont {Albanese}, \citenamefont
  {Ambrosino}, \citenamefont {Bachmann}, \citenamefont {Brezinsek},
  \citenamefont {Donne}, \citenamefont {Eich}, \citenamefont {Falchetto},
  \citenamefont {Federici} \emph {et~al.}}]{Turnyanskiy}%
  \BibitemOpen
  \bibfield  {author} {\bibinfo {author} {\bibfnamefont {M.}~\bibnamefont
  {Turnyanskiy}}, \bibinfo {author} {\bibfnamefont {R.}~\bibnamefont {Neu}},
  \bibinfo {author} {\bibfnamefont {R.}~\bibnamefont {Albanese}}, \bibinfo
  {author} {\bibfnamefont {R.}~\bibnamefont {Ambrosino}}, \bibinfo {author}
  {\bibfnamefont {C.}~\bibnamefont {Bachmann}}, \bibinfo {author}
  {\bibfnamefont {S.}~\bibnamefont {Brezinsek}}, \bibinfo {author}
  {\bibfnamefont {T.}~\bibnamefont {Donne}}, \bibinfo {author} {\bibfnamefont
  {T.}~\bibnamefont {Eich}}, \bibinfo {author} {\bibfnamefont {G.}~\bibnamefont
  {Falchetto}}, \bibinfo {author} {\bibfnamefont {G.}~\bibnamefont {Federici}},
   \emph {et~al.},\ }\href {\doibase
  http://dx.doi.org/10.1016/j.fusengdes.2015.04.041} {\bibfield  {journal}
  {\bibinfo  {journal} {Fusion Eng. and Design}\ }\textbf {\bibinfo {volume}
  {96-97}},\ \bibinfo {pages} {361} (\bibinfo {year} {2015})},\ \bibinfo {note}
  {{Proceedings} of the 28th Symposium On Fusion Technology
  (SOFT-28)}\BibitemShut {NoStop}%
\bibitem [{\citenamefont {Jones}\ \emph {et~al.}(2016)\citenamefont {Jones},
  \citenamefont {Thomas}, \citenamefont {Amendt}, \citenamefont {Hall},
  \citenamefont {Izumi}, \citenamefont {{Barrios~Garcia}}, \citenamefont
  {Hopkins}, \citenamefont {Chen}, \citenamefont {Dewald}, \citenamefont
  {Hinkel} \emph {et~al.}}]{Jones}%
  \BibitemOpen
  \bibfield  {author} {\bibinfo {author} {\bibfnamefont {O.~S.}\ \bibnamefont
  {Jones}}, \bibinfo {author} {\bibfnamefont {C.~A.}\ \bibnamefont {Thomas}},
  \bibinfo {author} {\bibfnamefont {P.~A.}\ \bibnamefont {Amendt}}, \bibinfo
  {author} {\bibfnamefont {G.~N.}\ \bibnamefont {Hall}}, \bibinfo {author}
  {\bibfnamefont {N.}~\bibnamefont {Izumi}}, \bibinfo {author} {\bibfnamefont
  {M.~A.}\ \bibnamefont {{Barrios~Garcia}}}, \bibinfo {author} {\bibfnamefont
  {L.~F.~B.}\ \bibnamefont {Hopkins}}, \bibinfo {author} {\bibfnamefont
  {H.}~\bibnamefont {Chen}}, \bibinfo {author} {\bibfnamefont {E.~L.}\
  \bibnamefont {Dewald}}, \bibinfo {author} {\bibfnamefont {D.~E.}\
  \bibnamefont {Hinkel}},  \emph {et~al.},\ }\href
  {http://stacks.iop.org/1742-6596/717/i=1/a=012026} {\bibfield  {journal}
  {\bibinfo  {journal} {J. Phys.: Conf. Series}\ }\textbf {\bibinfo {volume}
  {717}},\ \bibinfo {pages} {012026} (\bibinfo {year} {2016})}\BibitemShut
  {NoStop}%
\bibitem [{\citenamefont {Fundamenski}(2005)}]{fundamenski}%
  \BibitemOpen
  \bibfield  {author} {\bibinfo {author} {\bibfnamefont {W.}~\bibnamefont
  {Fundamenski}},\ }\href@noop {} {\bibfield  {journal} {\bibinfo  {journal}
  {Plasma Phys. Control. Fus.}\ }\textbf {\bibinfo {volume} {47}},\ \bibinfo
  {pages} {R163} (\bibinfo {year} {2005})}\BibitemShut {NoStop}%
\bibitem [{\citenamefont {Taitano}, \citenamefont {Chac{\'{o}}n},\ and\
  \citenamefont {Simakov}(2016)}]{IFP0D2V}%
  \BibitemOpen
  \bibfield  {author} {\bibinfo {author} {\bibfnamefont {W.~T.}\ \bibnamefont
  {Taitano}}, \bibinfo {author} {\bibfnamefont {L.}~\bibnamefont
  {Chac{\'{o}}n}}, \ and\ \bibinfo {author} {\bibfnamefont {A.~N.}\
  \bibnamefont {Simakov}},\ }\href {\doibase
  http://dx.doi.org/10.1016/j.jcp.2016.03.071} {\bibfield  {journal} {\bibinfo
  {journal} {Journal of Computational Physics}\ }\textbf {\bibinfo {volume}
  {318}},\ \bibinfo {pages} {391 } (\bibinfo {year} {2016})}\BibitemShut
  {NoStop}%
\bibitem [{\citenamefont {Kolobov}\ and\ \citenamefont
  {Arslanbekov}(2006)}]{VFPNeutrals}%
  \BibitemOpen
  \bibfield  {author} {\bibinfo {author} {\bibfnamefont {V.~I.}\ \bibnamefont
  {Kolobov}}\ and\ \bibinfo {author} {\bibfnamefont {R.~R.}\ \bibnamefont
  {Arslanbekov}},\ }\href {\doibase 10.1109/TPS.2006.875850} {\bibfield
  {journal} {\bibinfo  {journal} {IEEE Transactions on Plasma Science}\
  }\textbf {\bibinfo {volume} {34}},\ \bibinfo {pages} {895} (\bibinfo {year}
  {2006})}\BibitemShut {NoStop}%
\bibitem [{\citenamefont {Chankin}, \citenamefont {Coster},\ and\ \citenamefont
  {Meisl}(2012)}]{KIPPDev}%
  \BibitemOpen
  \bibfield  {author} {\bibinfo {author} {\bibfnamefont {A.~V.}\ \bibnamefont
  {Chankin}}, \bibinfo {author} {\bibfnamefont {D.~P.}\ \bibnamefont {Coster}},
  \ and\ \bibinfo {author} {\bibfnamefont {G.}~\bibnamefont {Meisl}},\ }\href
  {\doibase 10.1002/ctpp.201210039} {\bibfield  {journal} {\bibinfo  {journal}
  {Contrib. Plasma Phys.}\ }\textbf {\bibinfo {volume} {52}},\ \bibinfo {pages}
  {500} (\bibinfo {year} {2012})}\BibitemShut {NoStop}%
\bibitem [{\citenamefont {Chankin}\ and\ \citenamefont
  {Coster}(2014)}]{KIPPBenchmarks}%
  \BibitemOpen
  \bibfield  {author} {\bibinfo {author} {\bibfnamefont {A.~V.}\ \bibnamefont
  {Chankin}}\ and\ \bibinfo {author} {\bibfnamefont {D.~P.}\ \bibnamefont
  {Coster}},\ }\href {\doibase 10.1002/ctpp.201410047} {\bibfield  {journal}
  {\bibinfo  {journal} {Contrib. Plasma Phys.}\ }\textbf {\bibinfo {volume}
  {54}},\ \bibinfo {pages} {493} (\bibinfo {year} {2014})}\BibitemShut
  {NoStop}%
\bibitem [{\citenamefont {Reiter}, \citenamefont {Baelmans},\ and\
  \citenamefont {B{\"{o}}rner}(2005)}]{B2Eirene}%
  \BibitemOpen
  \bibfield  {author} {\bibinfo {author} {\bibfnamefont {D.}~\bibnamefont
  {Reiter}}, \bibinfo {author} {\bibfnamefont {M.}~\bibnamefont {Baelmans}}, \
  and\ \bibinfo {author} {\bibfnamefont {P.}~\bibnamefont {B{\"{o}}rner}},\
  }\href {\doibase 10.13182/FST47-172} {\bibfield  {journal} {\bibinfo
  {journal} {Fusion Science and Technology}\ }\textbf {\bibinfo {volume}
  {47}},\ \bibinfo {pages} {172} (\bibinfo {year} {2005})}\BibitemShut
  {NoStop}%
\bibitem [{\citenamefont {Thomas}\ \emph {et~al.}(2012)\citenamefont {Thomas},
  \citenamefont {Tzoufras}, \citenamefont {Robinson}, \citenamefont {Kingham},
  \citenamefont {Ridgers}, \citenamefont {Sherlock},\ and\ \citenamefont
  {Bell}}]{VFPReview}%
  \BibitemOpen
  \bibfield  {author} {\bibinfo {author} {\bibfnamefont {A.~G.~R.}\
  \bibnamefont {Thomas}}, \bibinfo {author} {\bibfnamefont {M.}~\bibnamefont
  {Tzoufras}}, \bibinfo {author} {\bibfnamefont {A.~P.~L.}\ \bibnamefont
  {Robinson}}, \bibinfo {author} {\bibfnamefont {R.~J.}\ \bibnamefont
  {Kingham}}, \bibinfo {author} {\bibfnamefont {C.~P.}\ \bibnamefont
  {Ridgers}}, \bibinfo {author} {\bibfnamefont {M.}~\bibnamefont {Sherlock}}, \
  and\ \bibinfo {author} {\bibfnamefont {A.~R.}\ \bibnamefont {Bell}},\ }\href
  {\doibase http://dx.doi.org/10.1016/j.jcp.2011.09.028} {\bibfield  {journal}
  {\bibinfo  {journal} {Journal of Computational Physics}\ }\textbf {\bibinfo
  {volume} {231}},\ \bibinfo {pages} {1051 } (\bibinfo {year} {2012})},\
  \bibinfo {note} {special Issue: Computational Plasma Physics}\BibitemShut
  {NoStop}%
\bibitem [{\citenamefont {Brown}\ and\ \citenamefont
  {Haines}(1997)}]{FermiDirac}%
  \BibitemOpen
  \bibfield  {author} {\bibinfo {author} {\bibfnamefont {S.~R.}\ \bibnamefont
  {Brown}}\ and\ \bibinfo {author} {\bibfnamefont {M.~G.}\ \bibnamefont
  {Haines}},\ }\href
  {https://www.cambridge.org/core/journals/journal-of-plasma-physics/article/transport-in-partially-degenerate-magnetized-plasmas-part-1-collision-operators/88FB03EFE0F8623BA6603ACF163B7D6A}
  {\bibfield  {journal} {\bibinfo  {journal} {Journal of Plasma Physics}\
  }\textbf {\bibinfo {volume} {58}},\ \bibinfo {pages} {577–600} (\bibinfo
  {year} {1997})}\BibitemShut {NoStop}%
\bibitem [{\citenamefont {Gregori}\ \emph {et~al.}(2004)\citenamefont
  {Gregori}, \citenamefont {Glenzer}, \citenamefont {Knight}, \citenamefont
  {Niemann}, \citenamefont {Price}, \citenamefont {Froula}, \citenamefont
  {Edwards}, \citenamefont {Town}, \citenamefont {Brantov}, \citenamefont
  {Rozmus},\ and\ \citenamefont {Bychenkov}}]{Gregori}%
  \BibitemOpen
  \bibfield  {author} {\bibinfo {author} {\bibfnamefont {G.}~\bibnamefont
  {Gregori}}, \bibinfo {author} {\bibfnamefont {S.~H.}\ \bibnamefont
  {Glenzer}}, \bibinfo {author} {\bibfnamefont {J.}~\bibnamefont {Knight}},
  \bibinfo {author} {\bibfnamefont {C.}~\bibnamefont {Niemann}}, \bibinfo
  {author} {\bibfnamefont {D.}~\bibnamefont {Price}}, \bibinfo {author}
  {\bibfnamefont {D.~H.}\ \bibnamefont {Froula}}, \bibinfo {author}
  {\bibfnamefont {M.~J.}\ \bibnamefont {Edwards}}, \bibinfo {author}
  {\bibfnamefont {R.~P.~J.}\ \bibnamefont {Town}}, \bibinfo {author}
  {\bibfnamefont {A.}~\bibnamefont {Brantov}}, \bibinfo {author} {\bibfnamefont
  {W.}~\bibnamefont {Rozmus}}, \ and\ \bibinfo {author} {\bibfnamefont
  {V.~{\relax Yu}.}\ \bibnamefont {Bychenkov}},\ }\href {\doibase
  10.1103/PhysRevLett.92.205006} {\bibfield  {journal} {\bibinfo  {journal}
  {Phys. Rev. Lett.}\ }\textbf {\bibinfo {volume} {92}},\ \bibinfo {pages}
  {205006} (\bibinfo {year} {2004})}\BibitemShut {NoStop}%
\bibitem [{\citenamefont {Luciani}, \citenamefont {Mora},\ and\ \citenamefont
  {Virmont}(1983)}]{LMV}%
  \BibitemOpen
  \bibfield  {author} {\bibinfo {author} {\bibfnamefont {J.~F.}\ \bibnamefont
  {Luciani}}, \bibinfo {author} {\bibfnamefont {P.}~\bibnamefont {Mora}}, \
  and\ \bibinfo {author} {\bibfnamefont {J.}~\bibnamefont {Virmont}},\ }\href
  {\doibase 10.1103/PhysRevLett.51.1664} {\bibfield  {journal} {\bibinfo
  {journal} {Phys. Rev. Lett.}\ }\textbf {\bibinfo {volume} {51}},\ \bibinfo
  {pages} {1664} (\bibinfo {year} {1983})}\BibitemShut {NoStop}%
\bibitem [{\citenamefont {Manheimer}, \citenamefont {Colombant},\ and\
  \citenamefont {Goncharov}(2008)}]{CMG}%
  \BibitemOpen
  \bibfield  {author} {\bibinfo {author} {\bibfnamefont {W.}~\bibnamefont
  {Manheimer}}, \bibinfo {author} {\bibfnamefont {D.}~\bibnamefont
  {Colombant}}, \ and\ \bibinfo {author} {\bibfnamefont {V.}~\bibnamefont
  {Goncharov}},\ }\href {\doibase 10.1063/1.2963078} {\bibfield  {journal}
  {\bibinfo  {journal} {Phys. Plasmas}\ }\textbf {\bibinfo {volume} {15}},\
  \bibinfo {pages} {083103} (\bibinfo {year} {2008})}\BibitemShut {NoStop}%
\bibitem [{\citenamefont {Batishchev}\ \emph {et~al.}(2002)\citenamefont
  {Batishchev}, \citenamefont {Bychenkov}, \citenamefont {Detering},
  \citenamefont {Rozmus}, \citenamefont {Sydora}, \citenamefont {Capjack},\
  and\ \citenamefont {Novikov}}]{Batischev2002}%
  \BibitemOpen
  \bibfield  {author} {\bibinfo {author} {\bibfnamefont {O.~V.}\ \bibnamefont
  {Batishchev}}, \bibinfo {author} {\bibfnamefont {V.~{\relax Yu}.}\
  \bibnamefont {Bychenkov}}, \bibinfo {author} {\bibfnamefont {F.}~\bibnamefont
  {Detering}}, \bibinfo {author} {\bibfnamefont {W.}~\bibnamefont {Rozmus}},
  \bibinfo {author} {\bibfnamefont {R.}~\bibnamefont {Sydora}}, \bibinfo
  {author} {\bibfnamefont {C.~E.}\ \bibnamefont {Capjack}}, \ and\ \bibinfo
  {author} {\bibfnamefont {V.~N.}\ \bibnamefont {Novikov}},\ }\href {\doibase
  10.1063/1.1461385} {\bibfield  {journal} {\bibinfo  {journal} {Phys.
  Plasmas}\ }\textbf {\bibinfo {volume} {9}},\ \bibinfo {pages} {2302}
  (\bibinfo {year} {2002})}\BibitemShut {NoStop}%
\bibitem [{\citenamefont {Brantov}\ and\ \citenamefont
  {Bychenkov}(2013)}]{Brantov2013}%
  \BibitemOpen
  \bibfield  {author} {\bibinfo {author} {\bibfnamefont {A.~V.}\ \bibnamefont
  {Brantov}}\ and\ \bibinfo {author} {\bibfnamefont {V.~{\relax Yu}.}\
  \bibnamefont {Bychenkov}},\ }\href {\doibase 10.1134/S1063780X13090018}
  {\bibfield  {journal} {\bibinfo  {journal} {Plasma Phys. Reports}\ }\textbf
  {\bibinfo {volume} {39}},\ \bibinfo {pages} {698} (\bibinfo {year}
  {2013})}\BibitemShut {NoStop}%
\bibitem [{\citenamefont {Hammett}\ and\ \citenamefont
  {Perkins}(1990)}]{Hammett}%
  \BibitemOpen
  \bibfield  {author} {\bibinfo {author} {\bibfnamefont {G.~W.}\ \bibnamefont
  {Hammett}}\ and\ \bibinfo {author} {\bibfnamefont {F.~W.}\ \bibnamefont
  {Perkins}},\ }\href {\doibase 10.1103/PhysRevLett.64.3019} {\bibfield
  {journal} {\bibinfo  {journal} {Phys. Rev. Lett.}\ }\textbf {\bibinfo
  {volume} {64}},\ \bibinfo {pages} {3019} (\bibinfo {year}
  {1990})}\BibitemShut {NoStop}%
\bibitem [{\citenamefont {Izacard}(2017)}]{Izacard2017}%
  \BibitemOpen
  \bibfield  {author} {\bibinfo {author} {\bibfnamefont {O.}~\bibnamefont
  {Izacard}},\ }\href {htpp://dx.doi.org/10.1017/S0022377817000150} {\bibfield
  {journal} {\bibinfo  {journal} {J. Plasma Phys.}\ }\textbf {\bibinfo {volume}
  {83}} (\bibinfo {year} {2017})}\BibitemShut {NoStop}%
\bibitem [{\citenamefont {Marocchino}\ \emph {et~al.}(2013)\citenamefont
  {Marocchino}, \citenamefont {Tzoufras}, \citenamefont {Atzeni}, \citenamefont
  {Schiavi}, \citenamefont {\relax{Ph} D~Nicola\"i}, \citenamefont {Mallet},
  \citenamefont {Tikhonchuk},\ and\ \citenamefont {\relax{J.-L.}
  Feugeas}}]{Marocchino}%
  \BibitemOpen
  \bibfield  {author} {\bibinfo {author} {\bibfnamefont {A.}~\bibnamefont
  {Marocchino}}, \bibinfo {author} {\bibfnamefont {M.}~\bibnamefont
  {Tzoufras}}, \bibinfo {author} {\bibfnamefont {S.}~\bibnamefont {Atzeni}},
  \bibinfo {author} {\bibfnamefont {A.}~\bibnamefont {Schiavi}}, \bibinfo
  {author} {\bibnamefont {\relax{Ph} D~Nicola\"i}}, \bibinfo {author}
  {\bibfnamefont {J.}~\bibnamefont {Mallet}}, \bibinfo {author} {\bibfnamefont
  {V.}~\bibnamefont {Tikhonchuk}}, \ and\ \bibinfo {author} {\bibnamefont
  {\relax{J.-L.} Feugeas}},\ }\href {\doibase 10.1063/1.4789878} {\bibfield
  {journal} {\bibinfo  {journal} {Phys. Plasmas}\ }\textbf {\bibinfo {volume}
  {20}},\ \bibinfo {pages} {022702} (\bibinfo {year} {2013})}\BibitemShut
  {NoStop}%
\bibitem [{\citenamefont {Kingham}\ and\ \citenamefont {Bell}(2004)}]{IMPACT}%
  \BibitemOpen
  \bibfield  {author} {\bibinfo {author} {\bibfnamefont {R.~J.}\ \bibnamefont
  {Kingham}}\ and\ \bibinfo {author} {\bibfnamefont {A.~R.}\ \bibnamefont
  {Bell}},\ }\href {\doibase https://doi.org/10.1016/j.jcp.2003.08.017}
  {\bibfield  {journal} {\bibinfo  {journal} {J. Comp. Phys.}\ }\textbf
  {\bibinfo {volume} {194}},\ \bibinfo {pages} {1} (\bibinfo {year}
  {2004})}\BibitemShut {NoStop}%
\bibitem [{\citenamefont {Epperlein}(1994)}]{Epperlein94}%
  \BibitemOpen
  \bibfield  {author} {\bibinfo {author} {\bibfnamefont {E.~M.}\ \bibnamefont
  {Epperlein}},\ }\href {\doibase 10.1063/1.870563} {\bibfield  {journal}
  {\bibinfo  {journal} {Phys. Plasmas}\ }\textbf {\bibinfo {volume} {1}},\
  \bibinfo {pages} {109} (\bibinfo {year} {1994})}\BibitemShut {NoStop}%
\bibitem [{\citenamefont {Mason}(1981)}]{implicitPoisson}%
  \BibitemOpen
  \bibfield  {author} {\bibinfo {author} {\bibfnamefont {R.~J.}\ \bibnamefont
  {Mason}},\ }\href {\doibase http://dx.doi.org/10.1016/0021-9991(81)90094-2}
  {\bibfield  {journal} {\bibinfo  {journal} {J. Comp. Phys.}\ }\textbf
  {\bibinfo {volume} {41}},\ \bibinfo {pages} {233 } (\bibinfo {year}
  {1981})}\BibitemShut {NoStop}%
\bibitem [{\citenamefont {Trubnikov}(1965)}]{Trubnikov}%
  \BibitemOpen
  \bibfield  {author} {\bibinfo {author} {\bibfnamefont {B.~A.}\ \bibnamefont
  {Trubnikov}},\ }\enquote {\bibinfo {title} {Particle interactions in a fully
  ionized plasma},}\ in\ \href@noop {} {\emph {\bibinfo {booktitle} {Review of
  Plasma Physics}}},\ Vol.~\bibinfo {volume} {1},\ \bibinfo {editor} {edited
  by\ \bibinfo {editor} {\bibfnamefont {A.~M.}\ \bibnamefont {Leontovich}}}\
  (\bibinfo  {publisher} {Consultants Bureau},\ \bibinfo {address} {New York},\
  \bibinfo {year} {1965})\ pp.\ \bibinfo {pages} {105--204}\BibitemShut
  {NoStop}%
\bibitem [{\citenamefont {Rosenbluth}, \citenamefont {MacDonald},\ and\
  \citenamefont {Judd}(1957)}]{Rosenbluth}%
  \BibitemOpen
  \bibfield  {author} {\bibinfo {author} {\bibfnamefont {M.~N.}\ \bibnamefont
  {Rosenbluth}}, \bibinfo {author} {\bibfnamefont {W.~M.}\ \bibnamefont
  {MacDonald}}, \ and\ \bibinfo {author} {\bibfnamefont {D.~L.}\ \bibnamefont
  {Judd}},\ }\href {\doibase 10.1103/PhysRev.107.1} {\bibfield  {journal}
  {\bibinfo  {journal} {Phys. Rev.}\ }\textbf {\bibinfo {volume} {107}},\
  \bibinfo {pages} {1} (\bibinfo {year} {1957})}\BibitemShut {NoStop}%
\bibitem [{\citenamefont {Meisl}, \citenamefont {Chankin},\ and\ \citenamefont
  {Coster}(2013)}]{KIPPRelax}%
  \BibitemOpen
  \bibfield  {author} {\bibinfo {author} {\bibfnamefont {G.}~\bibnamefont
  {Meisl}}, \bibinfo {author} {\bibfnamefont {A.~V.}\ \bibnamefont {Chankin}},
  \ and\ \bibinfo {author} {\bibfnamefont {D.~P.}\ \bibnamefont {Coster}},\
  }\href {\doibase http://dx.doi.org/10.1016/j.jnucmat.2013.01.064} {\bibfield
  {journal} {\bibinfo  {journal} {J. Nucl. Mater.}\ }\textbf {\bibinfo {volume}
  {438}},\ \bibinfo {pages} {S342 } (\bibinfo {year} {2013})},\ \bibinfo {note}
  {{\relax Proceedings} of the 20th International Conference on Plasma-Surface
  Interactions in Controlled Fusion Devices}\BibitemShut {NoStop}%
\bibitem [{\citenamefont {Xiong}\ \emph {et~al.}(2008)\citenamefont {Xiong},
  \citenamefont {Cohen}, \citenamefont {Rognlien},\ and\ \citenamefont
  {Xu}}]{Xiong}%
  \BibitemOpen
  \bibfield  {author} {\bibinfo {author} {\bibfnamefont {Z.}~\bibnamefont
  {Xiong}}, \bibinfo {author} {\bibfnamefont {R.~H.}\ \bibnamefont {Cohen}},
  \bibinfo {author} {\bibfnamefont {T.~D.}\ \bibnamefont {Rognlien}}, \ and\
  \bibinfo {author} {\bibfnamefont {X.~Q.}\ \bibnamefont {Xu}},\ }\href
  {\doibase http://dx.doi.org/10.1016/j.jcp.2008.04.004} {\bibfield  {journal}
  {\bibinfo  {journal} {Journal of Computational Physics}\ }\textbf {\bibinfo
  {volume} {227}},\ \bibinfo {pages} {7192 } (\bibinfo {year}
  {2008})}\BibitemShut {NoStop}%
\bibitem [{\citenamefont {Shoucri}\ and\ \citenamefont
  {Gagne}(1978)}]{opSplit}%
  \BibitemOpen
  \bibfield  {author} {\bibinfo {author} {\bibfnamefont {M.~M.}\ \bibnamefont
  {Shoucri}}\ and\ \bibinfo {author} {\bibfnamefont {R.~R.~J.}\ \bibnamefont
  {Gagne}},\ }\href {\doibase http://dx.doi.org/10.1016/0021-9991(78)90013-X}
  {\bibfield  {journal} {\bibinfo  {journal} {J. Comput. Phys.}\ }\textbf
  {\bibinfo {volume} {27}},\ \bibinfo {pages} {315} (\bibinfo {year}
  {1978})}\BibitemShut {NoStop}%
\bibitem [{\citenamefont {Cheng}(2010)}]{ShoucriBook}%
  \BibitemOpen
  \bibfield  {author} {\bibinfo {author} {\bibfnamefont {C.~Z.}\ \bibnamefont
  {Cheng}},\ }\enquote {\bibinfo {title} {Splitting methods for the
  {Vlasov-Maxwell} equations in plasma simulations},}\ in\ \href@noop {} {\emph
  {\bibinfo {booktitle} {Eulerian Codes for the Numerical Solution of the
  Kinetic Equations of Plasmas}}},\ \bibinfo {editor} {edited by\ \bibinfo
  {editor} {\bibfnamefont {M.}~\bibnamefont {Shoucri}}}\ (\bibinfo  {publisher}
  {Nova Science Publishers Inc.},\ \bibinfo {address} {Hauppauge NY},\ \bibinfo
  {year} {2010})\ pp.\ \bibinfo {pages} {1--21}\BibitemShut {NoStop}%
\bibitem [{\citenamefont {Matte}\ and\ \citenamefont
  {Virmont}(1982)}]{MatteVirmont}%
  \BibitemOpen
  \bibfield  {author} {\bibinfo {author} {\bibfnamefont {J.~P.}\ \bibnamefont
  {Matte}}\ and\ \bibinfo {author} {\bibfnamefont {J.}~\bibnamefont
  {Virmont}},\ }\href {\doibase 10.1103/PhysRevLett.49.1936} {\bibfield
  {journal} {\bibinfo  {journal} {Phys. Rev. Lett.}\ }\textbf {\bibinfo
  {volume} {49}},\ \bibinfo {pages} {1936} (\bibinfo {year}
  {1982})}\BibitemShut {NoStop}%
\bibitem [{\citenamefont {Beer}\ and\ \citenamefont {Hammett}(1996)}]{H2}%
  \BibitemOpen
  \bibfield  {author} {\bibinfo {author} {\bibfnamefont {M.~A.}\ \bibnamefont
  {Beer}}\ and\ \bibinfo {author} {\bibfnamefont {G.~W.}\ \bibnamefont
  {Hammett}},\ }\href {\doibase 10.1063/1.871538} {\bibfield  {journal}
  {\bibinfo  {journal} {Phys. Plasmas}\ }\textbf {\bibinfo {volume} {3}},\
  \bibinfo {pages} {4046} (\bibinfo {year} {1996})}\BibitemShut {NoStop}%
\bibitem [{\citenamefont {Snyder}, \citenamefont {Hammett},\ and\ \citenamefont
  {Dorland}(1997)}]{H3}%
  \BibitemOpen
  \bibfield  {author} {\bibinfo {author} {\bibfnamefont {P.~B.}\ \bibnamefont
  {Snyder}}, \bibinfo {author} {\bibfnamefont {G.~W.}\ \bibnamefont {Hammett}},
  \ and\ \bibinfo {author} {\bibfnamefont {W.}~\bibnamefont {Dorland}},\ }\href
  {\doibase 10.1063/1.872517} {\bibfield  {journal} {\bibinfo  {journal} {Phys.
  Plasmas}\ }\textbf {\bibinfo {volume} {4}},\ \bibinfo {pages} {3974}
  (\bibinfo {year} {1997})}\BibitemShut {NoStop}%
\bibitem [{\citenamefont {Breil}\ and\ \citenamefont {Maire}(2007)}]{CHIC}%
  \BibitemOpen
  \bibfield  {author} {\bibinfo {author} {\bibfnamefont {J.}~\bibnamefont
  {Breil}}\ and\ \bibinfo {author} {\bibfnamefont {{\relax P.-H}.}~\bibnamefont
  {Maire}},\ }\href {\doibase http://dx.doi.org/10.1016/j.jcp.2006.10.025}
  {\bibfield  {journal} {\bibinfo  {journal} {J. Comp. Phys.}\ }\textbf
  {\bibinfo {volume} {224}},\ \bibinfo {pages} {785} (\bibinfo {year}
  {2007})}\BibitemShut {NoStop}%
\bibitem [{\citenamefont {Albritton}\ \emph {et~al.}(1986)\citenamefont
  {Albritton}, \citenamefont {Williams}, \citenamefont {Bernstein},\ and\
  \citenamefont {Swartz}}]{AWBS}%
  \BibitemOpen
  \bibfield  {author} {\bibinfo {author} {\bibfnamefont {J.~R.}\ \bibnamefont
  {Albritton}}, \bibinfo {author} {\bibfnamefont {E.~A.}\ \bibnamefont
  {Williams}}, \bibinfo {author} {\bibfnamefont {I.~B.}\ \bibnamefont
  {Bernstein}}, \ and\ \bibinfo {author} {\bibfnamefont {K.~P.}\ \bibnamefont
  {Swartz}},\ }\href {\doibase 10.1103/PhysRevLett.57.1887} {\bibfield
  {journal} {\bibinfo  {journal} {Phys. Rev. Lett.}\ }\textbf {\bibinfo
  {volume} {57}},\ \bibinfo {pages} {1887} (\bibinfo {year}
  {1986})}\BibitemShut {NoStop}%
\bibitem [{\citenamefont {Bychenkov}\ \emph {et~al.}(1994)\citenamefont
  {Bychenkov}, \citenamefont {Myatt}, \citenamefont {Rozmus},\ and\
  \citenamefont {Tikhonchuk}}]{Bychenkov}%
  \BibitemOpen
  \bibfield  {author} {\bibinfo {author} {\bibfnamefont {V.~{\relax Yu}.}\
  \bibnamefont {Bychenkov}}, \bibinfo {author} {\bibfnamefont {J.}~\bibnamefont
  {Myatt}}, \bibinfo {author} {\bibfnamefont {W.}~\bibnamefont {Rozmus}}, \
  and\ \bibinfo {author} {\bibfnamefont {V.~T.}\ \bibnamefont {Tikhonchuk}},\
  }\href {\doibase 10.1103/PhysRevE.50.5134} {\bibfield  {journal} {\bibinfo
  {journal} {Phys. Rev. E}\ }\textbf {\bibinfo {volume} {50}},\ \bibinfo
  {pages} {5134} (\bibinfo {year} {1994})}\BibitemShut {NoStop}%
\bibitem [{\citenamefont {Bychenkov}\ \emph {et~al.}(1995)\citenamefont
  {Bychenkov}, \citenamefont {Myatt}, \citenamefont {Rozmus},\ and\
  \citenamefont {Tikhonchuk}}]{Bychenkov95}%
  \BibitemOpen
  \bibfield  {author} {\bibinfo {author} {\bibfnamefont {V.~{\relax Yu}.}\
  \bibnamefont {Bychenkov}}, \bibinfo {author} {\bibfnamefont {J.}~\bibnamefont
  {Myatt}}, \bibinfo {author} {\bibfnamefont {W.}~\bibnamefont {Rozmus}}, \
  and\ \bibinfo {author} {\bibfnamefont {V.~T.}\ \bibnamefont {Tikhonchuk}},\
  }\href {\doibase 10.1103/PhysRevE.52.6759} {\bibfield  {journal} {\bibinfo
  {journal} {Phys. Rev. E}\ }\textbf {\bibinfo {volume} {52}},\ \bibinfo
  {pages} {6759} (\bibinfo {year} {1995})}\BibitemShut {NoStop}%
\bibitem [{\citenamefont {Chang}\ and\ \citenamefont
  {Callen}(1992)}]{ChangCallen}%
  \BibitemOpen
  \bibfield  {author} {\bibinfo {author} {\bibfnamefont {Z.}~\bibnamefont
  {Chang}}\ and\ \bibinfo {author} {\bibfnamefont {J.~D.}\ \bibnamefont
  {Callen}},\ }\href {\doibase 10.1063/1.860125} {\bibfield  {journal}
  {\bibinfo  {journal} {Phys. Fluids B}\ }\textbf {\bibinfo {volume} {4}},\
  \bibinfo {pages} {1167} (\bibinfo {year} {1992})}\BibitemShut {NoStop}%
\bibitem [{\citenamefont {Sanmartin}\ \emph {et~al.}(1992)\citenamefont
  {Sanmartin}, \citenamefont {Ram\'{i}rez}, \citenamefont
  {Fern\'{a}ndez-Feria},\ and\ \citenamefont {Minotti}}]{Sanmartin}%
  \BibitemOpen
  \bibfield  {author} {\bibinfo {author} {\bibfnamefont {J.~R.}\ \bibnamefont
  {Sanmartin}}, \bibinfo {author} {\bibfnamefont {J.}~\bibnamefont
  {Ram\'{i}rez}}, \bibinfo {author} {\bibfnamefont {R.}~\bibnamefont
  {Fern\'{a}ndez-Feria}}, \ and\ \bibinfo {author} {\bibfnamefont
  {F.}~\bibnamefont {Minotti}},\ }\href {\doibase 10.1063/1.860366} {\bibfield
  {journal} {\bibinfo  {journal} {Phys. Fluids B}\ }\textbf {\bibinfo {volume}
  {4}},\ \bibinfo {pages} {3579} (\bibinfo {year} {1992})}\BibitemShut
  {NoStop}%
\bibitem [{\citenamefont {Ji}\ and\ \citenamefont {Held}(2014)}]{JiClosure}%
  \BibitemOpen
  \bibfield  {author} {\bibinfo {author} {\bibfnamefont {J.~Y.}\ \bibnamefont
  {Ji}}\ and\ \bibinfo {author} {\bibfnamefont {E.~D.}\ \bibnamefont {Held}},\
  }\href {\doibase 10.1063/1.4904906} {\bibfield  {journal} {\bibinfo
  {journal} {Phys. Plasmas}\ }\textbf {\bibinfo {volume} {21}},\ \bibinfo
  {pages} {122116} (\bibinfo {year} {2014})}\BibitemShut {NoStop}%
\bibitem [{\citenamefont {Joseph}\ and\ \citenamefont {Dimits}(2016)}]{Joseph}%
  \BibitemOpen
  \bibfield  {author} {\bibinfo {author} {\bibfnamefont {I.}~\bibnamefont
  {Joseph}}\ and\ \bibinfo {author} {\bibfnamefont {A.~M.}\ \bibnamefont
  {Dimits}},\ }\href {\doibase 10.1002/ctpp.201610043} {\bibfield  {journal}
  {\bibinfo  {journal} {Contrib. Plasma Phys.}\ ,\ \bibinfo {pages} {504}}
  (\bibinfo {year} {2016})}\BibitemShut {NoStop}%
\bibitem [{\citenamefont {Kaufman}(1975)}]{varpro}%
  \BibitemOpen
  \bibfield  {author} {\bibinfo {author} {\bibfnamefont {L.}~\bibnamefont
  {Kaufman}},\ }\href {\doibase 10.1007/BF01932995} {\bibfield  {journal}
  {\bibinfo  {journal} {BIT Numer. Math.}\ }\textbf {\bibinfo {volume} {15}},\
  \bibinfo {pages} {49} (\bibinfo {year} {1975})}\BibitemShut {NoStop}%
\bibitem [{\citenamefont {Turrell}, \citenamefont {Sherlock},\ and\
  \citenamefont {Rose}(2015)}]{Turrell}%
  \BibitemOpen
  \bibfield  {author} {\bibinfo {author} {\bibfnamefont {A.}~\bibnamefont
  {Turrell}}, \bibinfo {author} {\bibfnamefont {M.}~\bibnamefont {Sherlock}}, \
  and\ \bibinfo {author} {\bibfnamefont {S.}~\bibnamefont {Rose}},\ }\href
  {\doibase http://dx.doi.org/10.1016/j.jcp.2015.06.034} {\bibfield  {journal}
  {\bibinfo  {journal} {J. Comp. Phys.}\ }\textbf {\bibinfo {volume} {299}},\
  \bibinfo {pages} {144 } (\bibinfo {year} {2015})}\BibitemShut {NoStop}%
\bibitem [{\citenamefont {Sherlock}, \citenamefont {Brodrick},\ and\
  \citenamefont {Ridgers}(2017)}]{SherlockSNB}%
  \BibitemOpen
  \bibfield  {author} {\bibinfo {author} {\bibfnamefont {M.}~\bibnamefont
  {Sherlock}}, \bibinfo {author} {\bibfnamefont {J.~P.}\ \bibnamefont
  {Brodrick}}, \ and\ \bibinfo {author} {\bibfnamefont {C.~P.}\ \bibnamefont
  {Ridgers}},\ }\href {http://dx.doi.org/10.1063/1.4986095} {\bibfield  {journal} 
  {\bibinfo  {journal} {Phys. Plasmas}\ }\textbf {\bibinfo {volume} {24}},\
  \bibinfo {pages} {082706} (\bibinfo {year} {2017})}\BibitemShut
  {NoStop}%
\bibitem [{\citenamefont {Dum}(1978{\natexlab{a}})}]{Dum1}%
  \BibitemOpen
  \bibfield  {author} {\bibinfo {author} {\bibfnamefont {C.~T.}\ \bibnamefont
  {Dum}},\ }\href {\doibase 10.1063/1.862338} {\bibfield  {journal} {\bibinfo
  {journal} {Phys. Fluids}\ }\textbf {\bibinfo {volume} {21}},\ \bibinfo
  {pages} {945} (\bibinfo {year} {1978}{\natexlab{a}})}\BibitemShut {NoStop}%
\bibitem [{\citenamefont {Dum}(1978{\natexlab{b}})}]{Dum2}%
  \BibitemOpen
  \bibfield  {author} {\bibinfo {author} {\bibfnamefont {C.~T.}\ \bibnamefont
  {Dum}},\ }\href {\doibase 10.1063/1.862339} {\bibfield  {journal} {\bibinfo
  {journal} {Phys. Fluids}\ }\textbf {\bibinfo {volume} {21}},\ \bibinfo
  {pages} {956} (\bibinfo {year} {1978}{\natexlab{b}})}\BibitemShut {NoStop}%
\bibitem [{\citenamefont {Langdon}(1980)}]{Langdon}%
  \BibitemOpen
  \bibfield  {author} {\bibinfo {author} {\bibfnamefont {A.~B.}\ \bibnamefont
  {Langdon}},\ }\href {\doibase 10.1103/PhysRevLett.44.575} {\bibfield
  {journal} {\bibinfo  {journal} {Phys. Rev. Lett.}\ }\textbf {\bibinfo
  {volume} {44}},\ \bibinfo {pages} {575} (\bibinfo {year} {1980})}\BibitemShut
  {NoStop}%
\bibitem [{\citenamefont {Matte}\ \emph {et~al.}(1988)\citenamefont {Matte},
  \citenamefont {Lamoureux}, \citenamefont {Moller}, \citenamefont {Yin},
  \citenamefont {Delettrez}, \citenamefont {Virmont},\ and\ \citenamefont
  {Johnston}}]{Matte}%
  \BibitemOpen
  \bibfield  {author} {\bibinfo {author} {\bibfnamefont {J.~P.}\ \bibnamefont
  {Matte}}, \bibinfo {author} {\bibfnamefont {M.}~\bibnamefont {Lamoureux}},
  \bibinfo {author} {\bibfnamefont {C.}~\bibnamefont {Moller}}, \bibinfo
  {author} {\bibfnamefont {R.~Y.}\ \bibnamefont {Yin}}, \bibinfo {author}
  {\bibfnamefont {J.}~\bibnamefont {Delettrez}}, \bibinfo {author}
  {\bibfnamefont {J.}~\bibnamefont {Virmont}}, \ and\ \bibinfo {author}
  {\bibfnamefont {T.~W.}\ \bibnamefont {Johnston}},\ }\href
  {http://stacks.iop.org/0741-3335/30/i=12/a=004} {\bibfield  {journal}
  {\bibinfo  {journal} {Plasma Phys. Control. Fusion}\ }\textbf {\bibinfo
  {volume} {30}},\ \bibinfo {pages} {1665} (\bibinfo {year}
  {1988})}\BibitemShut {NoStop}%
\bibitem [{\citenamefont {Ridgers}\ \emph {et~al.}(2008)\citenamefont
  {Ridgers}, \citenamefont {Thomas}, \citenamefont {Kingham},\ and\
  \citenamefont {Robinson}}]{InvBrem}%
  \BibitemOpen
  \bibfield  {author} {\bibinfo {author} {\bibfnamefont {C.~P.}\ \bibnamefont
  {Ridgers}}, \bibinfo {author} {\bibfnamefont {A.~G.~R.}\ \bibnamefont
  {Thomas}}, \bibinfo {author} {\bibfnamefont {R.~J.}\ \bibnamefont {Kingham}},
  \ and\ \bibinfo {author} {\bibfnamefont {A.~P.~L.}\ \bibnamefont
  {Robinson}},\ }\href {\doibase 10.1103/PhysRevLett.100.075003} {\bibfield
  {journal} {\bibinfo  {journal} {Phys. Plasmas}\ }\textbf {\bibinfo {volume}
  {100}},\ \bibinfo {eid} {092311} (\bibinfo {year} {2008})}\BibitemShut
  {NoStop}%
\bibitem [{\citenamefont {Ridgers}, \citenamefont {Kingham},\ and\
  \citenamefont {Thomas}(2008)}]{reemerge}%
  \BibitemOpen
  \bibfield  {author} {\bibinfo {author} {\bibfnamefont {C.~P.}\ \bibnamefont
  {Ridgers}}, \bibinfo {author} {\bibfnamefont {R.~J.}\ \bibnamefont
  {Kingham}}, \ and\ \bibinfo {author} {\bibfnamefont {A.~G.~R.}\ \bibnamefont
  {Thomas}},\ }\href {\doibase 10.1103/PhysRevLett.100.075003} {\bibfield
  {journal} {\bibinfo  {journal} {Phys. Rev. Lett.}\ }\textbf {\bibinfo
  {volume} {100}},\ \bibinfo {pages} {075003} (\bibinfo {year}
  {2008})}\BibitemShut {NoStop}%
\bibitem [{\citenamefont {Joglekar}\ \emph {et~al.}(2016)\citenamefont
  {Joglekar}, \citenamefont {Ridgers}, \citenamefont {Kingham},\ and\
  \citenamefont {Thomas}}]{Archis}%
  \BibitemOpen
  \bibfield  {author} {\bibinfo {author} {\bibfnamefont {A.~S.}\ \bibnamefont
  {Joglekar}}, \bibinfo {author} {\bibfnamefont {C.~P.}\ \bibnamefont
  {Ridgers}}, \bibinfo {author} {\bibfnamefont {R.~J.}\ \bibnamefont
  {Kingham}}, \ and\ \bibinfo {author} {\bibfnamefont {A.~G.~R.}\ \bibnamefont
  {Thomas}},\ }\href {\doibase 10.1103/PhysRevE.93.043206} {\bibfield
  {journal} {\bibinfo  {journal} {Phys. Rev. E}\ }\textbf {\bibinfo {volume}
  {93}},\ \bibinfo {pages} {043206} (\bibinfo {year} {2016})}\BibitemShut
  {NoStop}%
\bibitem [{\citenamefont {Montgomery}\ \emph {et~al.}(2015)\citenamefont
  {Montgomery}, \citenamefont {Albright}, \citenamefont {Barnak}, \citenamefont
  {Chang}, \citenamefont {Davies}, \citenamefont {Fiksel}, \citenamefont
  {Froula}, \citenamefont {Kline}, \citenamefont {MacDonald}, \citenamefont
  {Sefkow}, \citenamefont {Yin},\ and\ \citenamefont {Betti}}]{Montgomery}%
  \BibitemOpen
  \bibfield  {author} {\bibinfo {author} {\bibfnamefont {D.~S.}\ \bibnamefont
  {Montgomery}}, \bibinfo {author} {\bibfnamefont {B.~J.}\ \bibnamefont
  {Albright}}, \bibinfo {author} {\bibfnamefont {D.~H.}\ \bibnamefont
  {Barnak}}, \bibinfo {author} {\bibfnamefont {P.~Y.}\ \bibnamefont {Chang}},
  \bibinfo {author} {\bibfnamefont {J.~R.}\ \bibnamefont {Davies}}, \bibinfo
  {author} {\bibfnamefont {G.}~\bibnamefont {Fiksel}}, \bibinfo {author}
  {\bibfnamefont {D.~H.}\ \bibnamefont {Froula}}, \bibinfo {author}
  {\bibfnamefont {J.~L.}\ \bibnamefont {Kline}}, \bibinfo {author}
  {\bibfnamefont {M.~J.}\ \bibnamefont {MacDonald}}, \bibinfo {author}
  {\bibfnamefont {A.~B.}\ \bibnamefont {Sefkow}}, \bibinfo {author}
  {\bibfnamefont {L.}~\bibnamefont {Yin}}, \ and\ \bibinfo {author}
  {\bibfnamefont {R.}~\bibnamefont {Betti}},\ }\href {\doibase
  10.1063/1.4906055} {\bibfield  {journal} {\bibinfo  {journal} {Phys.
  Plasmas}\ }\textbf {\bibinfo {volume} {22}},\ \bibinfo {eid} {010703}
  (\bibinfo {year} {2015})}\BibitemShut {NoStop}%
\bibitem [{\citenamefont {Strozzi}\ \emph {et~al.}(2015)\citenamefont
  {Strozzi}, \citenamefont {Perkins}, \citenamefont {Marinak}, \citenamefont
  {Larson}, \citenamefont {Koning},\ and\ \citenamefont {Logan}}]{Strozzi2015}%
  \BibitemOpen
  \bibfield  {author} {\bibinfo {author} {\bibfnamefont {D.~J.}\ \bibnamefont
  {Strozzi}}, \bibinfo {author} {\bibfnamefont {L.~J.}\ \bibnamefont
  {Perkins}}, \bibinfo {author} {\bibfnamefont {M.~M.}\ \bibnamefont
  {Marinak}}, \bibinfo {author} {\bibfnamefont {D.~J.}\ \bibnamefont {Larson}},
  \bibinfo {author} {\bibfnamefont {J.~M.}\ \bibnamefont {Koning}}, \ and\
  \bibinfo {author} {\bibfnamefont {B.~G.}\ \bibnamefont {Logan}},\ }\href
  {\doibase 10.1017/S0022377815001348} {\bibfield  {journal} {\bibinfo
  {journal} {J. Plasma Phys.}\ }\textbf {\bibinfo {volume} {81}} (\bibinfo
  {year} {2015}),\ 10.1017/S0022377815001348}\BibitemShut {NoStop}%
\bibitem [{\citenamefont {Izacard}\ \emph {et~al.}(2016)\citenamefont
  {Izacard}, \citenamefont {Holland}, \citenamefont {James},\ and\
  \citenamefont {Brennan}}]{Islands}%
  \BibitemOpen
  \bibfield  {author} {\bibinfo {author} {\bibfnamefont {O.}~\bibnamefont
  {Izacard}}, \bibinfo {author} {\bibfnamefont {C.}~\bibnamefont {Holland}},
  \bibinfo {author} {\bibfnamefont {S.~D.}\ \bibnamefont {James}}, \ and\
  \bibinfo {author} {\bibfnamefont {D.~P.}\ \bibnamefont {Brennan}},\ }\href
  {\doibase http://dx.doi.org/10.1063/1.4941704} {\bibfield  {journal}
  {\bibinfo  {journal} {Phys. Plasmas}\ }\textbf {\bibinfo {volume} {23}},\
  \bibinfo {pages} {022304} (\bibinfo {year} {2016})}\BibitemShut {NoStop}%
\bibitem [{\citenamefont {Brantov}\ \emph {et~al.}(1996)\citenamefont
  {Brantov}, \citenamefont {Bychenkov}, \citenamefont {Tikhonchuk},\ and\
  \citenamefont {Rozmus}}]{BrantovNoB}%
  \BibitemOpen
  \bibfield  {author} {\bibinfo {author} {\bibfnamefont {A.~V.}\ \bibnamefont
  {Brantov}}, \bibinfo {author} {\bibfnamefont {V.~{\relax Yu}.}\ \bibnamefont
  {Bychenkov}}, \bibinfo {author} {\bibfnamefont {V.~T.}\ \bibnamefont
  {Tikhonchuk}}, \ and\ \bibinfo {author} {\bibfnamefont {W.}~\bibnamefont
  {Rozmus}},\ }\href@noop {} {\bibfield  {journal} {\bibinfo  {journal} {JETP}\
  }\textbf {\bibinfo {volume} {42}},\ \bibinfo {pages} {716} (\bibinfo {year}
  {1996})}\BibitemShut {NoStop}%
\end{thebibliography}
\end{document}